\documentclass[journal=jacsat,manuscript=article]{achemso}

\usepackage[version=3]{mhchem} 
\usepackage{hyperref}
\usepackage{graphicx}
\usepackage{amsmath}
\usepackage{amssymb}
\usepackage{float} 
\usepackage{subfiles}
\usepackage{xcolor}
\usepackage{xfrac}
\usepackage{epstopdf}
\usepackage{pdfpages}
\AppendGraphicsExtensions{.tif}

\author{Diptabrata Paul}
\email{diptabrata.paul@students.iiserpune.ac.in}
\altaffiliation{These authors contributed equally to this work}
\author{Rahul Chand}
\altaffiliation{These authors contributed equally to this work}
\author{G V Pavan Kumar}
\affiliation[IISER]{Department of Physics, Indian Institute of Science Education and Research (IISER), Pune 411008, India}
\email{pavan@iiserpune.ac.in}
\title{Optothermal evolution of active colloidal matter in defocused laser trap}  

\begin{document}


\begin{abstract}
Optothermal interaction of active colloidal matter can facilitate environmental cues which can influence the dynamics of active soft matter systems. The optically induced thermal effect can be harnessed to study non-equilibrium thermodynamics as well as applied to self-propel colloids and form assemblies. In this work, we employ a defocused laser trap to form self-evolving colloidal active matter. The optothermal interaction of the active colloids in both focused and defocused optical trap has been investigated to ascertain their thermophoretic behavior, which shows a long-range attraction and a short-range repulsion between the colloids. The optical gradient field enabled attraction and the short-range repulsion between the active colloids have been harnessed to form re-configurable dynamic assembly. Additionally, the assembly undergoes self-evolution as a new colloid joins the structure. Further, we show that the incident polarization state of the optical field can be employed as a parameter to modulate the structural orientation of the active colloids. The simple defocused optical field-enabled assembly can serve as a model to understand the collective dynamics of active matter systems, and can be harnessed as re-configurable microscopic engine.
\end{abstract}

\section{Introduction}

Ever since the pioneering work of Ashkin \cite{ashkin1970,ashkin1986}, the optical forces facilitated by optical tweezers have been extensively utilized to trap and manipulate colloidal matter as well as to investigate their dynamics \cite{volpe_spr2006, volpe_trq2006, righini2007, garcia2006, kall2010, chen2011, partha2014, ganapathy2016, ganapathy2021, grier2003, dholakia2010, feldmann2014, trapping_dholakia2016}. In addition to optical potentials, the tweezer platform can also be harnessed for localized heating \cite{baffou2010, stefano2011, kall2018, baffou_book2017, baffou2020}. The resulting optothermal interactions have been employed to trap and manipulate colloids through thermophoresis \cite{duhr2006, braun2013, piazza2008, vandana2020}, thermo-osmosis \cite{wurger2013, thermo_osmosis_cichos2016, hydrodynamic_cichos2022} and thermoelectric effects \cite{thermo_electric_yuebing2018, tiwari2021} as well as to understand the non-equilibrium nature of the interaction \cite{wurger2010}. In this context, active matter has become an important system of study specifically because it can self-propel, absorbing energy from the environment and can mimic systems which are out of thermodynamic equilibrium \cite{ramaswami2010, sano2010, ramaswamy2013, nedev2015,falko2021, martin2021}. Additionally, naturally occurring active matter systems like swimming bacteria \cite{swinney2010,cates2012} and flocks of birds \cite{ballerini2008, vicsek2012} provide a test bed for many biophysical and cellular phenomena. The wealth of information in these systems motivated investigation of artificial active matter systems such as active colloids with optical, acoustic or chemical potentials \cite{dreyfus2005, clemens2016, falko2018, martin2021, takatori2016, theurkauff2012} as well as electric \cite{Janus_efield} and magnetic fields \cite{ambarish2009}.

Interaction of an artificial active colloid with an optical field leads to heating and consequent release of energy, which further creates a thermal gradient \cite{sano2010, utsab2018, falko2018, martin2021}. The temperature distribution acts as an environmental cue to influence the dynamics of active colloids leading to thermophoresis \cite{wurger2007,piazza2008}. The effect can be harnessed for self-thermophoretic propulsion \cite{martin2021}, feedback-controlled ordering of colloids \cite{utsab2018} as well as for studying microscopic engines \cite{falko2018}. As such, generating and studying the dynamic assembly of colloidal matter is particularly interesting because these can lead to better understanding of the collective dynamics and act as a model for various biological systems \cite{clemens2016, volpe_prx2016, clemens2020}. To that end, previous approaches of generating spatio-temporally ordered assembly involved holographic optical trapping as well as multiplexing of the incident optical beams to trap multiple colloids \cite{grier1998,curtis2002,grier2003,grier2006HOT,damet_hydro2012,utsab2018,Franzl2021}. However, a single optical beam enabled spatio-temporal ordering of active colloids remains to be explored in detail.

Motivated by that, herein we report spatio-temporal ordering of iron oxide infused polystyrene active colloids (PS ACs) in a defocused laser trap. The colloids undergo self-thermophoretic motion under asymmetric laser illumination, rendering their active nature. Additionally, the generated thermal gradient acts as an environmental cue for the colloids in the vicinity. This leads to thermophoretic motion of the PS ACs towards the heated region and results in a long-range attraction and a short-range repulsion between a trapped and a migrating colloid. Upon encountering the defocused optical field, the counteracting effect of attraction due to optical gradient potential and short-range repulsion due to thermal gradient, the colloids form ordered assembly as shown in the schematic Fig. \ref{f1}(a) and timeseires Fig. \ref{f1}(d). Additionally, we show that the polarization of the incident beam can dynamically modulate the structural orientation of the assembly of PS AC. Depending on the extent of defocusing, higher number of colloids can be trapped to form an ordered assembly.

\section{Results and discussion}

\begin{figure}
\includegraphics[width = 450 pt]{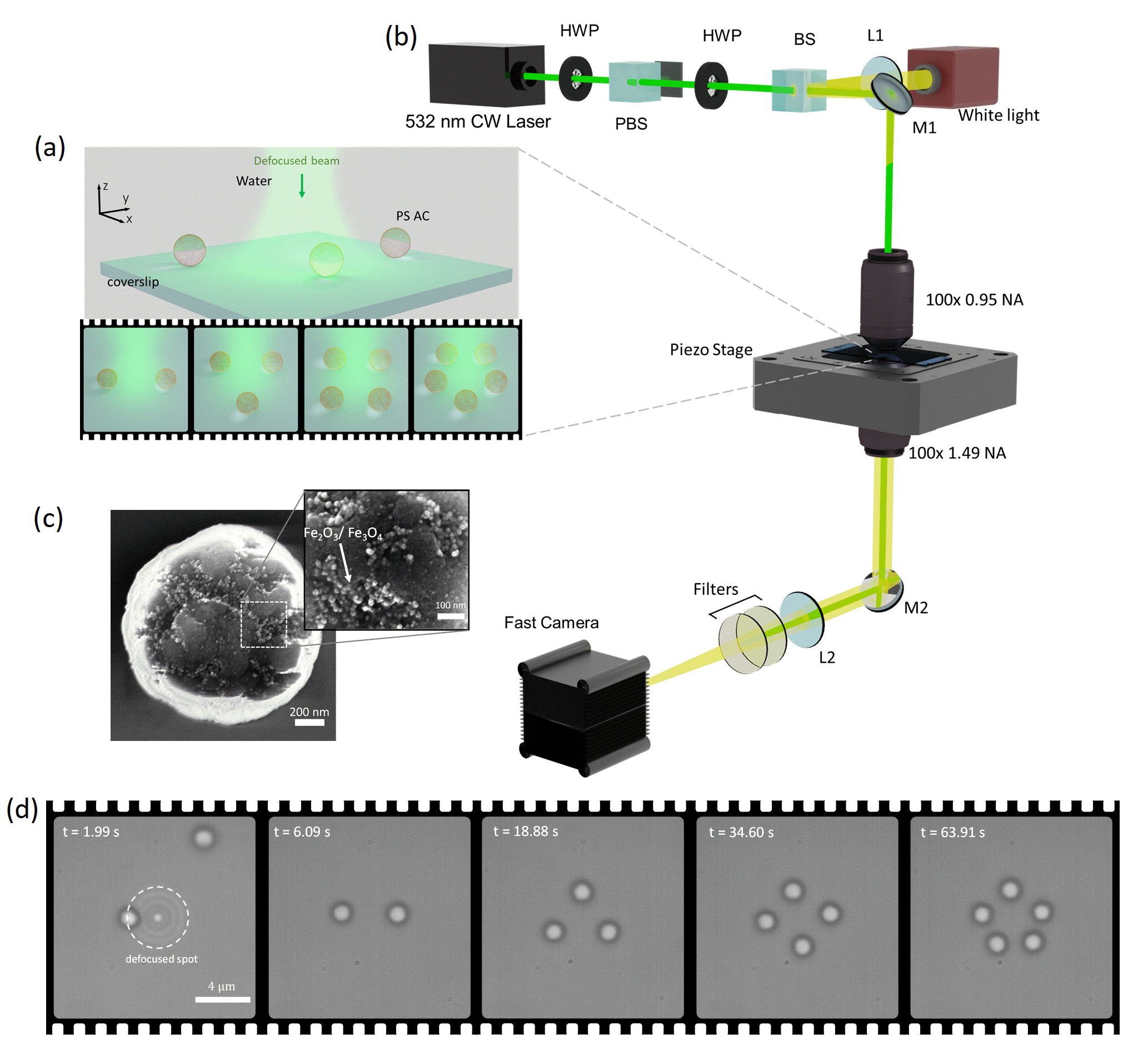}
\caption{\label{f1} Optothermal assembly of reconfigurable colloidal matter. (a) Polystyrene active colloids (PS ACs, diameter $1.31$ $\mu$m) infused with $\textrm{Fe}_2\textrm{O}_3/\textrm{Fe}_3\textrm{O}_4$ $\sim15$ nm nanoparticles with $\sim30\%$ mass fraction, are dispersed in water and placed in a microchamber enclosed by two glass coverslips. The nanoparticles are heated by incident laser beam and lead to thermal gradient in the surrounding medium, causing thermophoretic migration of the nearby PS ACs and dynamic assembly. (b) A dual channel optical microscope setup is used for the experiments. A $100\times$ $0.95$ NA objective lens is used for focusing/defocusing of incident laser at wavelength $\lambda= 532$ nm and a $100\times$ $1.49$ NA lens is used to collect the signal and projected onto the Fast camera ($100-1000$ fps) using relay optics. (c) Scanning electron micrograph of a PS AC shows the embedded $\textrm{Fe}_2\textrm{O}_3/\textrm{Fe}_3\textrm{O}_4$ nanoparticles. (d) Timeseries of the assembly of PS ACs under a defocused illumination.}
\end{figure}

The colloids used in the experiments are composed of polystyrene having diameter $1.31$ $\mu$m with nanoparticulate (diameter $\sim 15$ nm) iron oxide ($\textrm{Fe}_2\textrm{O}_3/\textrm{Fe}_3\textrm{O}_4$) distributed ($\sim 30\%$ mass fraction) throughout the pores within the polymer colloid interior, rendering them symmetric (Microparticles GmbH) (Fig. \ref{f1}(c) and supporting information S1). Each nanoparticle act like a point heat source and collectively they lead to asymmetric temperature distribution due to absorption of asymmetrically incident laser beam, resulting in self-thermophoresis, rendering the colloids active (Ultraviolet-visible absorption spectra shown in supporting information S1). The polystyrene active colloids (PS ACs) are dispersed in deionized water within a microchamber (height $\sim 100$ $\mu$m) enclosed with two glass coverslips and placed on a piezo stage. A dual channel optical microscope setup is used for the experiments with a $\lambda=532$ nm laser beam incident from the top side with a $100\times$ $0.95$ NA lens as shown in Fig. 1(b). The signal is collected from the bottom side with an oil immersion $100\times$ $1.49$ NA objective lens. Continuous variation of the laser power is performed by using a combination of half-wave plate (HWP) and polarizing beam-splitter (PBS). The excitation wavelength from the collected signal is rejected by using a combination of notch and edge filter and the signal is projected to a fast camera (fps $100-1000$) to record the dynamics of the PS ACs. The trajectories of the dynamics are extracted by employing Trackmate \cite{trackmate}. The experimental design allows us to independently modulate the focusing conditions of the excitation beam from the collection path.

In absence of any environmental perturbation, the PS ACs undergo unbiased Brownian motion. The diffusion constant for the freely diffusing PS ACs is $\textrm{D}=0.14$ $\mu \textrm{m}^2\textrm{s}^{-1}$. This is lower than the bulk diffusion coefficient ($\textrm{D}_0=0.33$ $\mu \textrm{m}^2\textrm{s}^{-1}$) and is due to their proximity to the glass surface (supporting information S2) \cite{brenner1961, volpe_book}. The freely diffusing PS ACs in the vicinity of an optically trapped and heated colloid undergo thermophoretic motion towards the heated region as a result of the generated temperature gradient \cite{piazza2008}. Additionally, the self-heating of the PS ACs undergoing thermophoretic motion leads to short-range repulsion between a trapped and an incoming colloid. Finally, under the collective influence of the defocused optical field and the generated thermal gradeint, the PS ACs attain a dynamic equilibrium to form a stable structural orientation as shown in Fig. 1(d) (SI Video1). Unlike the active colloids, passive colloids ($1.01$ $\mu$m polystyrene colloids in aqueous medium) do not exhibit any active dynamics in the defocused optical field and get trapped to form a two-dimensional array (SI Video 2). Thus, to understand the phenomena, it is important to first investigate the dynamic characteristics of individual PS AC in an focused optical trap.

\subsection{Self-thermophoresis of the colloids}

\begin{figure}[t]
\includegraphics[width = 450 pt]{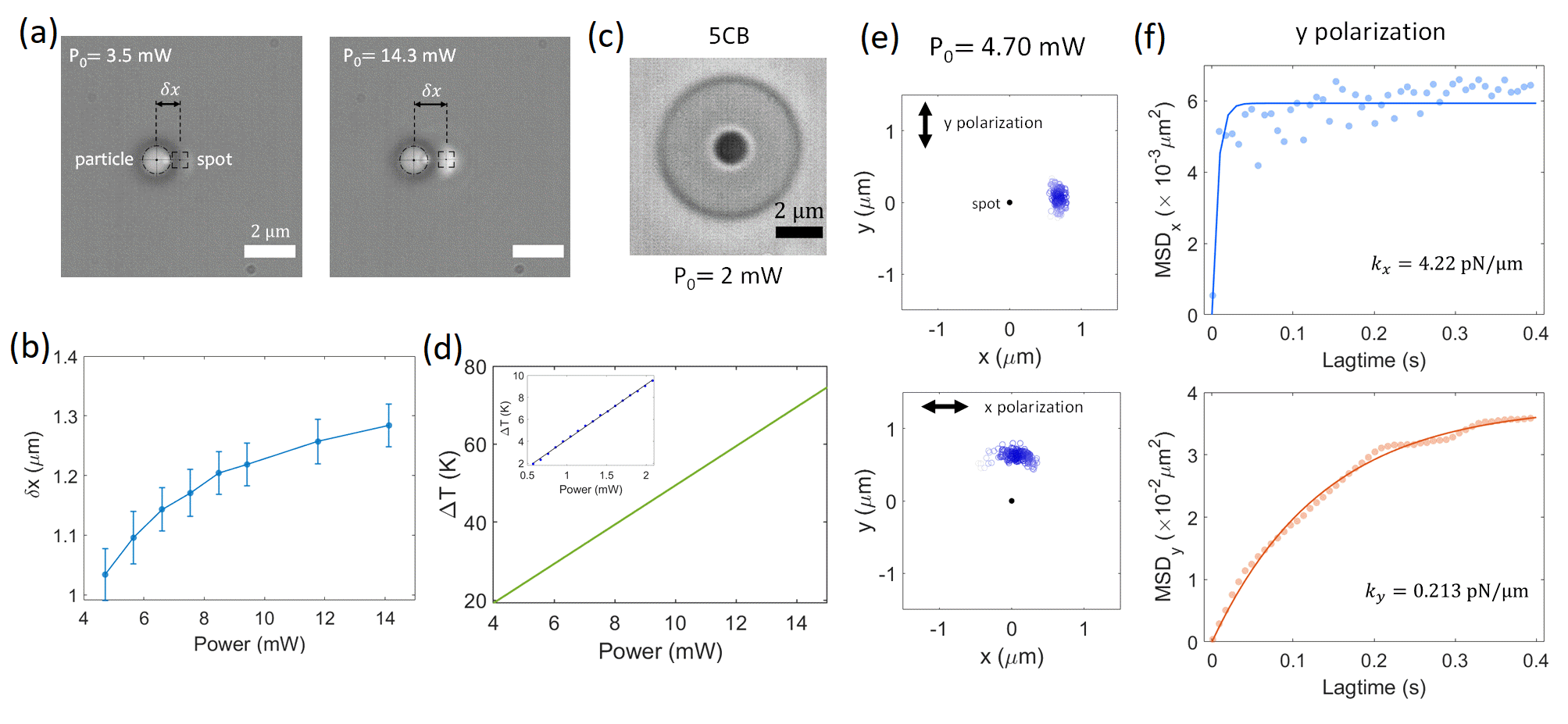}
\caption{\label{f2} Self-thermophoretic motion of a PS AC in the optical trap. (a) Under focused illumination, the trapped PS AC heats up and undergoes self-thermophoresis resulting from thermo-osmotic slipping of fluid along the colloid surface. The colloid settles off center under the influence of harmonic restoring optical force and thermally induced drift force away from the focal spot. (b) The equilibrium position of the trapped PS AC from the focal spot increases non-linearly with the laser power. (c) A bright-field optical image of liquid crystal phase transition from nematic to isotropic around an immobilised PS AC centrally heated with a focused laser power $\textrm{P}_0= 2.0$ mW. (d) Temperature increment of the PS AC as a function of laser power is estimated from the radius of the isotropic phase region. Temperature increment at higher laser power can be estimated by extrapolating the measured temperature increment at lower power shown in the inset. (e) Polarization of the incident beam modulates the direction of slipping of the colloid $-$ the colloid moves along $x$ direction for $y$ polarized beam and along $y$ direction for $x$ polarized beam. (f) The corresponding spring constant ($k_x,k_y$) for $y$ polarized incident beam is obtained by calculating corresponding mean square displacement in $x$ ($\textrm{MSD}_x$) and $y$ ($\textrm{MSD}_y$) direction and is more along $x$ direction compared to the $y$ direction.}
\end{figure}

A focused laser beam leads to generation of gradient optical force which can trap a colloid in a fluidic environment. The gradient optical force and the consequent trap stiffness for the colloids can be determined by tracking the position distribution of the colloid in a very low power optical trap (supporting information S3) \cite{volpe_book}. Upon increasing the power, the large optical field at the focus also leads to heating of the PS ACs \cite{baffou_book2017}. The elliptical nature of the incident optical field (supporting information S4) along with the motion of the colloid in the trap leads to asymmetric illumination and consequent thermal gradient on the colloid surface. This leads to an interfacial flow of the fluid, confined to a thin layer along the colloid surface, termed as thermo-osmotic slip flow \cite{thermo_osmosis_cichos2016} and consequent self-thermophoretic motion of the colloid, recently investigated for Janus colloids as well as symmetric microswimmers \cite{wurger2013,utsab2018,Franzl2021}. The colloid settles in an off-center position ($\delta x$) where the harmonic optical force ($\mathrm{F_{opt}} = k\delta x $, $k$ is trap stiffness) pulling the colloid towards the center balances the thermal gradient induced drift force ($\mathrm{F_{drift}} = \gamma v = -\gamma \mathrm{D_T \nabla T}$, $\mathrm{D_T}$ is thermo-diffusion coefficient, $\gamma$ represent bulk viscous coefficient) resulting from the fluid flow, as shown in \ref{f2}(a). The distance from the focus center ($\delta x$) at which this dynamic equilibrium is reached increases non-linearly with laser power, shown in Fig. \ref{f2}(b) (SI Video 3 and supporting information S5). The non-linearity has been investigated by considering temperature dependence of the thermo-diffusion coefficient ($\mathrm{D_T}$) as well as the Soret coefficient ($\mathrm{S_T}$) (supporting information S5) \cite{Iacopini2006, braibanti_piazza_thermophoresis, bechinger_thermophoresis}. Further analysis and quantification of the associated forces and approximate drift velocity has been investigated and shown in supporting information S6.

Experimental insight into the approximate maximum surface temperature at a given laser power is estimated by studying the nematic to isotropic phase transition of 5CB liquid crystal around a centrally heated immobile PS AC (supporting information S7) as shown in Fig. \ref{f2}(c), which scales linearly with incident laser power (Fig. \ref{f2}(d)) \cite{horn1978refractive,martin2021}. Additionally, the direction towards which a colloid will undergo self-thermophoretic drift depends on the polarization of the incident laser beam (SI video 4). For a $y$ polarized focused beam, the trapped PS AC drifts along $x$ direction (\ref{f2}(e)) and vice-versa for $x$ polarized focused beam (\ref{f2}(e)). This effect can be attributed to the elliptic shape of the electric field intensity distribution at the focus, with major axis along the linear polarization direction (supporting information S4), contributing to the heating. Thus, a colloid can drift along the minor axis ($x$ direction for $y$ polarized beam and vice versa for $x$ polarized beam), which will lead to lesser heating at a given distance from the potential minima compared to the same distance along major axis and hence the forces can be balanced relatively easily. The corresponding trap stiffness is obtained by calculating mean square displacements (MSD) and fitting it with $\mathrm{MSD}(\tau) = 2k_B T/k[1-e^{-|\tau|/\tau_{ot}}]$, where $\tau$ is the lag time and $\tau_{ot} = \gamma/k$ is trap characteristic time \cite{volpe_book}. Fig. \ref{f2}(f) shows the MSD along $x$ ($\mathrm{MSD}_x$) and $y$ ($\mathrm{MSD}_y$) direction for $y$ polarized incident beam. The corresponding trap stiffness along $x$ direction $k_x = 4.22$ $\mathrm{pN/\mu m}$ is one order magnitude more than that in the $y$ direction $k_y = 0.213$ $\mathrm{pN/\mu m}$. This is due to the fact that the colloid drifts in $x$ direction, which in this case represents the radial direction, and consequently is pulled towards the center along this axis due to the optical gradient force ($\mathrm{F_{opt}}= k(\mathrm{P_0})\cdot\delta x \approx 4.9$ $\mathrm{pN}$) and drifts due to the self-thermophoretic effect. This polarization dependence of the self-thermophoretic drifting direction opens up an additional parameter for manipulation of propulsion direction of active colloids. Further, the temperature distribution set up by a trapped and heated colloid results in an environmental cue for the motion of the freely diffusing PS ACs leading to unconventional interaction between them, which is discussed below.

\subsection{Thermophoretic hovering of the colloids}

\begin{figure}[t]
\includegraphics[width = 450 pt]{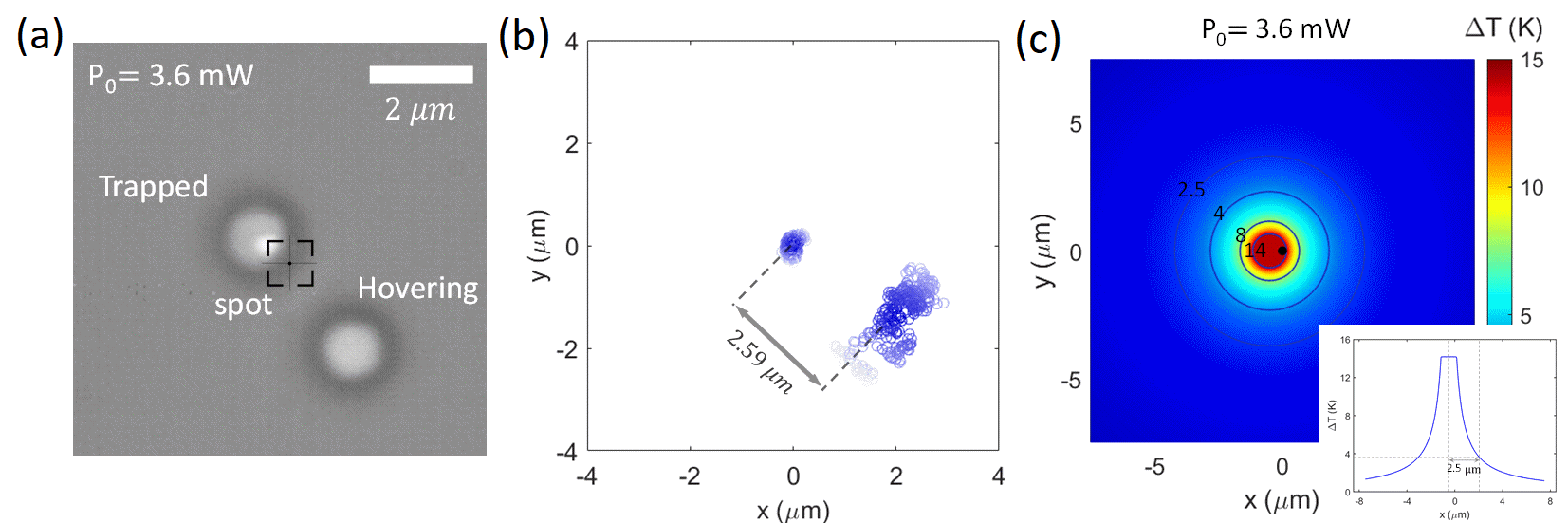}
\caption{\label{f3} Thermophoretic hovering of PS AC. (a) A heated trapped colloid at the focal spot results in temperature distribution in the surrounding medium. A second PS AC undergoes thermophoretic migration towards the heat center and eventually hovers near a trapped colloid at a certain distance. (b) Position distribution of the trapped and hovering colloid. (c) The temperature distribution that can be generated due to off-center trapped ($\delta x = -0.5$ $\mu$m)colloid at power $\mathrm{P_0}=3.6$ mW can be estimated from the experimentally measured temperature using 5CB, which assumes uniform heating. The corresponding line profile along $x$ axis is shown in the inset.}
\end{figure}

The environmental cue generated due to a trapped and heated colloid perturbs the Brownian motion of the PS ACs in the surrounding region resulting in their directed motion towards the heat center due to the thermophoresis \cite{piazza2008}. The experimental results indicate that a second colloid migrates towards the heat center due to thermophoresis and eventually near to the trapped colloid undergoes hovering at a certain distance, shown in fig. \ref{f3}(a). The corresponding position distributions of the trapped and hovering PS ACs are shown in fig. \ref{f3}(b), exhibit that at laser power $\mathrm{P_0}=3.6$ mW, the hovering occurs at distance $\sim 2.59$ $\mu$m. The temperature increment $\mathrm{\Delta T^{est}}$ for the trapped colloid is estimated by $\mathrm{\Delta T^{est} = \Delta T^{exp}}e^{-\delta x^2/2w_0^2}$, which assumes uniform heating of the off-center ($\delta x = -0.5$ $\mu$m) trapped colloid. The corresponding temperature distribution due to the trapped colloid will approximately vary as, $\textrm{T}(r)=\Delta \textrm{T}\sfrac{a}{r} +\textrm{T}_0$, where $r$ is the distance and $a$ is the radius of the colloid, $\Delta \textrm{T}$ is the temperature increment on the surface of the trapped colloid, $\textrm{T}_0= 298$ K is the ambient temperature\cite{thermo_osmosis_cichos2016} (supporting information S8). The generated radially symmetric temperature distribution is shown in fig. \ref{f3}(c). Inset shows the line profile of the temperature increment along $x$ axis.

Migrating towards a heated region indicates that far from the heat center the colloids have negative Soret coefficient ($\mathrm{S_T}$). However, the hovering resulting from thermophoretic repulsion can be attributed to temperature dependence of $\mathrm{S_T}$, i.e., assuming $\mathrm{S_T}=\mathrm{S^{\infty}_{T}}(1-e^{(T^*-T)/T_{f}})$ \cite{bechinger_thermophoresis,braibanti_piazza_thermophoresis}. $T^*$ represents the temperature at which $\mathrm{S_T}$ inverts sign and $T_f$ is a fitting parameter. For the experimental data, the temperature increment at the region where the hovering occurs ($d\approx 2.5$ $\mu$m) is $\mathrm{\Delta T}\approx 4$ K for $\mathrm{\Delta T_{max} \approx 14.2}$ K of the trapped colloid, as shown in fig. \ref{f3}(c), implying $T^*\approx 302$ K, similar to the value of $T^*$ for various colloids \cite{bechinger_thermophoresis,braibanti_piazza_thermophoresis}. Closer to heat center from that region $\mathrm{S_T}$ may invert its sign, resulting in inversion of thermophoretic velocity $v=-\textrm{D}_\textrm{T} \nabla \textrm{T}$ ($\mathrm{D_T = S_T D_0}$ is thermo-diffusion coefficient). Alternatively, encountering some scattered light near to a trapped colloid, the incoming colloid may attain a temperature $\textrm{T} \geq \textrm{T}(r)$ due to its absorption. Consequently, the migrating PS AC no longer encounters a positive thermal gradient ($\nabla \textrm{T}$) and become zero or switches its sign near to this position. Hence the thermophoretic velocity $v=-\textrm{D}_\textrm{T} \nabla \textrm{T}$ may have a value close to $0$ or invert its sign around this region. Any motion away or towards the trapped colloid will render fluctuation of the thermophoretic velocity sign, leading to the hovering of the second colloid.

The hovering distance depends on the temperature distribution due to the first colloid and hence can be modulated by changing the laser power (SI Video 5, supporting information S9). Thus, a heated colloid leads to a long-range attraction acting up to distance $10$-$12$ $\mu$m due to thermophoretic migration other colloids and a short-range repulsion acting in the range $1$-$5$ $\mu$m depending on the incident laser power, leading to their thermophoretic hovering. Such activity induced remote modulation of motion behaviour is significantly different from that of passive dielectric colloids, and creates an avenue for understanding as well as modification of collective dynamics of active matter systems \cite{vicsek2012}.

\subsection{Dynamics of colloidal pair in defocused optical trap}

\begin{figure}[t]
\includegraphics[width = 460 pt]{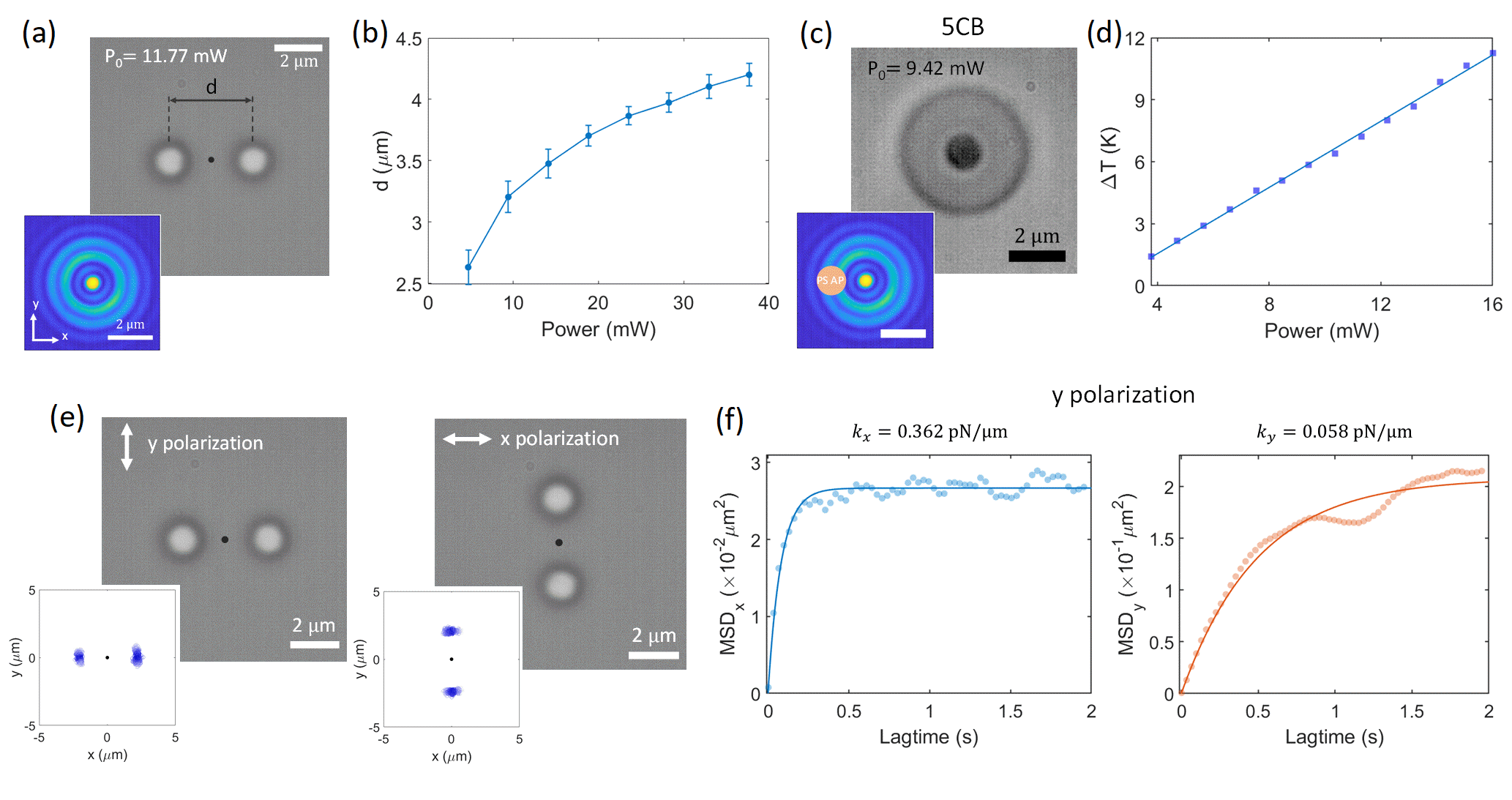}
\caption{\label{f4} Dynamics of colloidal pair in defocused optical trap. (a) Under defocused illumination, the optical field forms multiple concentric Airy rings. Two PS ACs can get trapped in this optical field, staying diametrically opposite to each other due to their thermal gradient. (b) For a fixed defocusing (diameter of defocus $\sim 4 $ $\mu$m), the distance between the colloid depends on the incident power and increases as the laser power is increased. (c) A bright-field optical image of  5CB liquid crystal phase transition from nematic to isotropic around a heated immobilised PS AC with a defocused laser power $\textrm{P}_0= 9.42$ mW. The position of the colloid with respect to the defocused spot is indicated in the inset. (d) Temperature increment of the PS AC as a function of laser power is estimated from the radius of the isotropic phase region. (e) The orientation of the PS ACs in the defocused spot additionally depends on the polarization of incident beam. The colloids stay along $x$ direction for $y$ polarized beam and stay along $y$ direction for $x$ polarized beam. (f) Calculated $\mathrm{MSD}_x$ and $\mathrm{MSD}_y$ and the corresponding trap-stiffness of the colloids for $y$ polarized incident beam.}
\end{figure}

While a focused beam leads to hovering of the colloids around a trapped PS AC, a defocused beam is advantageous for generating an optical field in which multiple colloids can get trapped with equal stiffness, and heat up to equal extent. Fig. \ref{f4}(a) shows assembly of two PS ACs trapped in diametrically opposite positions in the defocused field shown in the inset. The black spot indicates the beam center. The defocused beam has a central high intensity part followed by concentric Airy rings forming a circular optical field region as shown in the inset \cite{novotny2012principles}. The qualitative spatial distribution of the gradient potential and the gradient optical force of such defocused optical field can be extracted by scanning an immobile PS AC in the $xy$ plane and extracting the temperature increment. Since both the temperature increment and the optical trapping potential is proportional to the intensity profile of the incident optical field, the measure give us the qualitative optical gradient potential profile (supporting information S10). Alternatively, quantitatively the gradient optical potential difference between the central high intensity part and the peripheral region can be obtained by trapping passive dielectric colloids in those regions and analysing the trap stiffness (SI Video 2, supporting information S10). By examining the dynamics of passive dielectric dielectric $1.01$ $\mu$m PS colloids in the defocused optical field, the minima of the central potential is obtained to be $\mathrm{U^{cent}_{min}} =-4.4$ $k_BT$ and that for the peripheral region is $\mathrm{U^{peri}_{min}} =-3.6$ $k_BT$ at laser power $\mathrm{P_0} = 14$ mW.

The diametrically opposite trapping position shown in fig. \ref{f4}(a) is attributed to the fact that each of the PS ACs get heated up to equal extent under the effect of optical field and hence leads to repulsion due to thermal field of each other. The interparticle distance depends on the extent of defocusing as well as the incident laser power. Fig. \ref{f4}(b) shows that for fixed defocusing, the interparticle distance ($d$) increases non-linearly with power. The non-linearity can be attributed to the self-thermophoretic motion of the colloids as well as their repulsion from each other. To find out the approximate maximum temperature of a PS AC under such defocused illumination we study the nematic to isotropic transition of 5CB liquid crystal around a heated colloid shown in Fig. \ref{f4}(c). The inset shows the approximate position of the PS AC at the defocused spot. The maximum temperature variation with power is shown in Fig. \ref{f4}(d). The maximum temperature increment attained due to incident laser power of $\approx 16$ mW does not exceed $\approx12$ K.

The orientation of the PS ACs under such defocused illumination can be further modulated by changing the linear polarization state of the incident beam as shown in Fig. \ref{f4}(e). The two PS ACs orient themselves along $x$ direction for $y$ polarized beam and vice-versa for $x$ polarized incident beam (SI Video 6). The inset shows the position distribution obtained over 8 seconds. The trapping state for $y$ polarized incident beam (laser power $\mathrm{P_0} = 11.77$ mW) is further characterized by calculating the MSD along $x$ ($\mathrm{MSD}_x$) and $y$ ($\mathrm{MSD}_y$) direction and obtaining the corresponding trap-stiffness as shown in fig. \ref{f4}(f). The obtained trap-stiffness along $x$ direction $k_x = 0.362$ $\mathrm{pN/\mu m}$ is one order of magnitude higher than that of along $y$ direction $k_y = 0.058$ $\mathrm{pN/\mu m}$. The result could be understood considering their diametrically opposite arrangement along $x$ axis and resulting thermophoretic repulsive force and gradient optical field force acts along $x$ axis. The polarization dependence of the orientation can be attributed to the electric field intensity distribution under such defocused illumination and the corresponding optical forces (supporting information S4 and S12). The normalized electric field intensity distribution is more along $x$ direction for $y$ polarized beam and more along $y$ direction for $x$ polarized beam. Numerically the gradient optical force on the active colloids have been estimated by modelling the colloids using Maxwell-Garnett effective medium theory \cite{maxwell_garnett1,maxwell_garnett2} (supporting information S11) and evaluating the force in the defocused optical fields (supporting information S12). The coupled motion of the PS ACs in the defocused laser beam is reminiscent of many biological active matters as well as of fundamental importance due to the synchronization of motion through environmental cues \cite{swinney2010}. The activity induced self-evolution is characteristic of all naturally occurring active matter systems and motivates the investigation of dynamic assembly of multiple active colloids.

\subsection{Evolution of active colloidal matter}

\begin{figure}
\includegraphics[width = 400 pt]{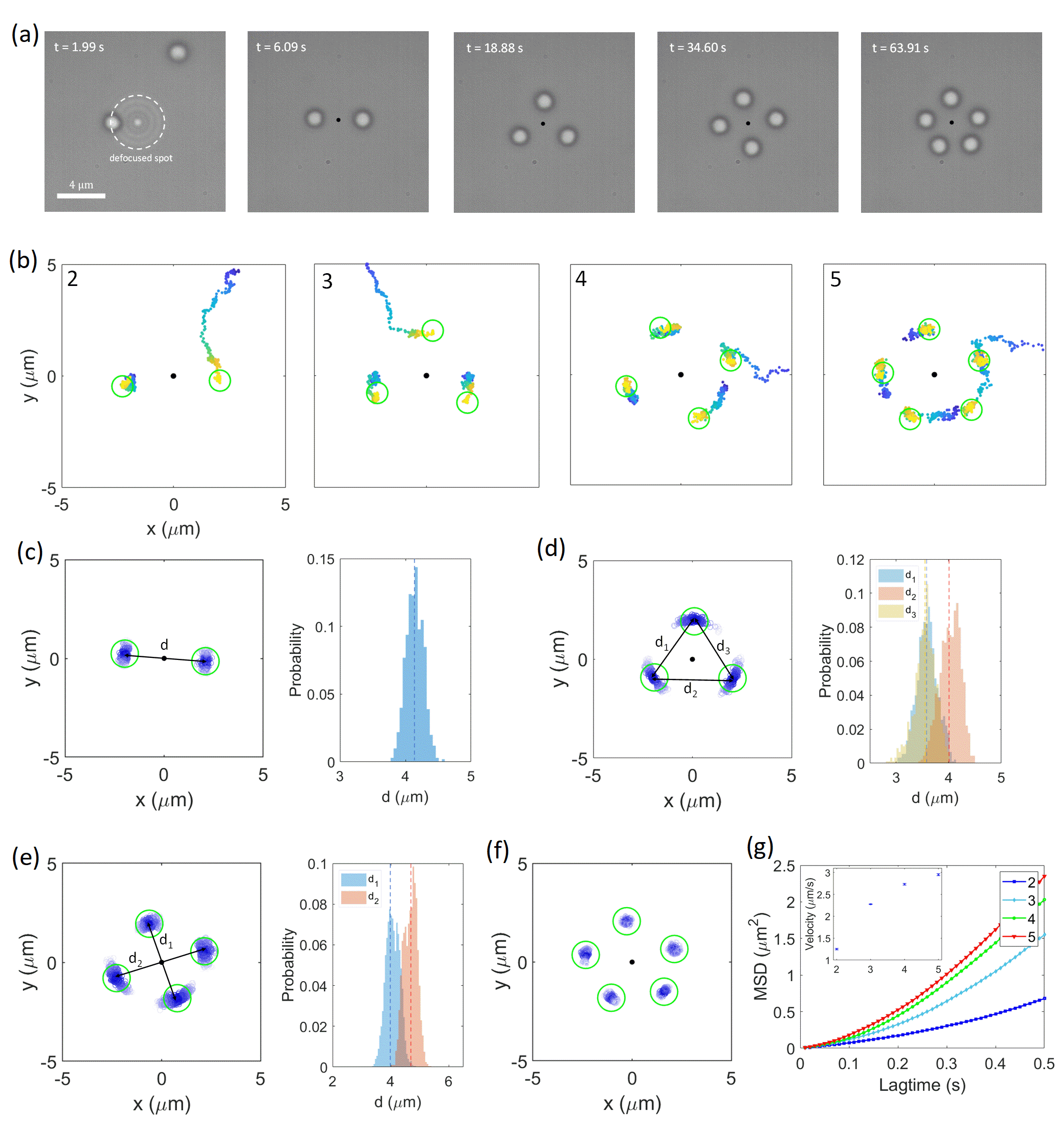}
\caption{\label{f5} Self-evolution of active colloidal matter under defocused illumination. (a) The time series of the assembly under a defocused illumination at $\textrm{P}_0 = 14.13$ mW. The trapping and heating of single PS AC ($\Delta\textrm{T}$ $\approx 10$ K) leads to a temperature distribution in the surrounding region and thermophoretic migration of other colloids towards the trapped heat center and formation of colloidal assembly. Black spot indicates the position of beam center (b) As a new colloid joins the assembly, the existing PS ACs re-position themselves in the assembly and forms a dynamic structure with the new colloid included. The dynamic equilibrium positions are indicated by the green open circles. (c) Two PS ACs settle in diametrically opposite position having average distance $d = 4.14$ $\mu$m, (d) Three in the vertices of a triangle with the distances between the colloids having distribution about $d_1=d_3\approx3.57$ $\mu$m and $d_2=4.03$ $\mu$m. (e) Four PS ACs form a diamond shaped quadrilateral geometry having diagonal distance distribution about $d_1=4.07$ $\mu$m and $d_2=4.69$ $\mu$m. (f) Five PS ACs form a pentagonal geometry. (g) The thermophoretic velocity of an incoming colloid is obtained by investigating the mean-squared displacement (MSD). The obtained velocity (given in the inset) of an incoming colloid show incremental trend.}
\end{figure}

Trapped PS ACs in the defocused laser spot generate a temperature distribution in the surrounding region (supporting information S8) and result in thermophoretic motion of other colloids in the vicinity. The described experimental arrangement can thus lead to self-evolution of the structures as multiple PS ACs migrate towards the spatio-temporally static optical field. The primary driving mechanism for this assembly is the thermophoretic behaviour  of the active colloids due to generated temperature gradient as well as the optical gradient force led spatio-temporal trapping. The optical gradient potential on the active colloids has been evaluated by extracting their position distribution. The gradient optical force experienced by the colloids along the radial direction is approximated by the corresponding trap stiffness $\approx0.447$ $\mathrm{pN/\mu m}$ at incident laser power $\mathrm{P_0}=14.13$ mW (supporting information S12). Fig. \ref{f5}(a) shows the dynamic assembly of multiple PS ACs as they migrate one by one towards the heated colloids in a $y$ polarized defocused beam (SI Video 1). The temporal self-evolution of the dynamically stable structures is shown in Fig. \ref{f5}(b). It can be seen that as a new colloid joins the structure, the existing colloidal arrangement re-organizes to form a stable assembly with the new colloid. In contrast, passive colloids (polystyrene $1.01$ $\mu$m) do not exhibit active motion under such illumination configuration and get trapped to form two dimensional array (SI Video 2).

The dynamic property of assembly of two PS ACs trapped in diametrically opposite position is quantified by their position distribution and their distance as shown in Fig. \ref{f5}(c). The distance between the colloids is shown in the histogram with average distance $\approx4.14$ $\mu$m. The assembly evolves into a triangle with three PS ACs as shown in Fig. \ref{f5}(d), with each colloid settling at the vertices. The distance between the colloids is indicated in the histogram, with two distributions having mean $\approx3.57$ $\mu$m and $\approx4.02$ $\mu$m. The larger distance between the horizontally oriented colloids is due to the electric field intensity distribution due to $y$ polarized beam. Four PS ACs reorganize themselves to form a diamond shaped quadrilateral structure as shown in Fig. \ref{f5}(e). The diagonal distance between the colloids exhibits two distributions with average $\approx4.07$ $\mu$m and $\approx4.68$ $\mu$m. The larger distance is between the colloids aligned horizontally and similar to the three PS AC case, it can be attributed to the polarization dependent electric field intensity distribution in the sample plane and the corresponding optical gradient forces (supporting information S4 and S12) . Five colloids lead to formation of a pentagonal structure as shown in Fig. \ref{f5}(f). Other than the effect of generated thermal gradient on the assembly process and their arrangement, we have also evaluated the effect of scattered light on neighbouring colloids and the resultant optical binding effect has been considered \cite{dholakia2010}. We have employed full wave three dimensional finite element method (FEM) simulation (COMSOL Multiphysics 5.1) and modelled the colloids as spherical particle with material properties evaluated using Maxwell-Garnett effective medium theory (supporting information S11) \cite{maxwell_garnett1,maxwell_garnett2}. The resultant scattering force is of the order $\approx 0.01$ pN, one order magnitude less than the gradient force on the colloids (supporting information S13), thus having a weaker effect on the assembly. It is important to note that the exact nature of the dynamic structure formation depends on the identical characteristic of the colloids, deformed or smaller/bigger sized colloids might lead to formation of skewed geometry in comparison to the ideal ones.

\begin{figure}
\includegraphics[width = 380 pt]{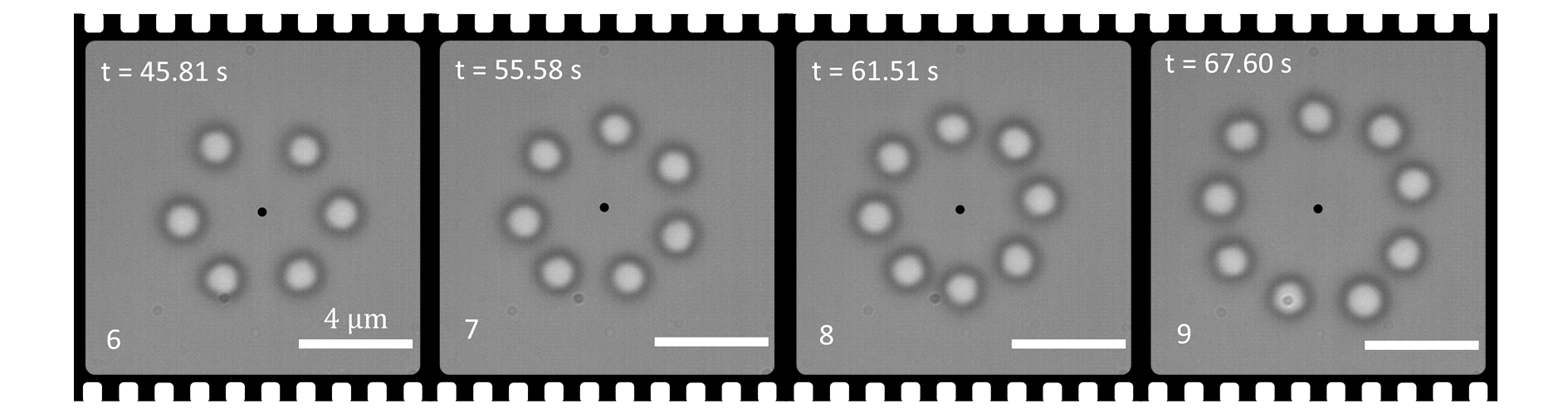}
\caption{\label{f6} Colloidal assembly with higher number of colloids. With increasing time, and for larger defocusing, dynamic assembly for 6-9 PS ACs can also be formed. The black spot indicates the beam center. (SI Video 7)}
\end{figure}

As the number of colloids in the assembly increase, the maximum temperature attained by the system increases (supporting information S8). Consequently, the velocity ($v$) of the directed motion with which a new colloid joins the assembly increases and can be found out by considering the mean squared displacement (MSD). The MSD for such a directed motion is given by $\textrm{MSD}(\tau)=v^2 \tau^2+ 4\textrm{D}\tau$, $\tau$ being the lag time \cite{sano2010}. Fig. \ref{f5}(g) shows the mean-squared displacement of the $n^\textrm{th}$ colloid as they undergo thermophoretic motion towards the assembly of $(n-1)$ colloid(s). The corresponding velocity ($v$) is shown in the inset and indicates an increasing trend having $v = 1.25$ $\mathrm{\mu m\,s^{-1}}$ for the second colloid to $v=2.94$ $\mathrm{\mu m\,s^{-1}}$ for the fifth colloid. 

The effect of thermal convection in the assembly process have been evaluated by employing FEM simulation. The simulation considers a 2D axisymmetric geometry and a centrally heated colloid, since a 3D simulation is computationally exhaustive. The numerical simulation indicates that a centrally heated colloid at laser power $\mathrm{P_0}=10$ mW leads to maximum surface temperature increment $\mathrm{\Delta T \approx 50}$ K, resulting in maximum thermal convective flow velocity $\approx 63$ $\mathrm{nm\, s^{-1}}$ (supporting information S14). An assembly of heated colloid may act like an extended heat source but for our experimental configuration even for an assembly of five colloids in the defocused beam at power $\mathrm{P_0}=14$ mW, the estimated temperature increment is $\approx 20$ K (supporting information S8). Thus, the effect of buoyancy driven thermal convection can be considered minimal with respect to the thermophoretic motion of the colloids. The corresponding evaluation of the thermo-diffusion coefficient ($\mathrm{D_T}$) and the Soret coefficient ($\mathrm{S_T}$) has been evaluated by ignoring the temperature dependence for simplicity. It has been calculated by considering the path ($r(t)$) of the second colloid due to the temperature distribution of the first colloid, which approximately varies as $r^{-1}$. The obtained value of Soret coefficient was found to be $\mathrm{S_T} = -115.16$ $\textrm{K}^{-1}$ (supporting information S15). The negative sign implies the migration towards a heated region.

Higher number of colloids can also form a stable colloidal matter through this process, as shown in Fig. \ref{f6} (SI Video 7). With a denser colloidal solution, the structure can evolve to form dynamic assembly of higher number of colloids in short time. However, it must be noted that the defocusing of the incident laser has to be increased for this purpose to ensure sufficient distance between two colloids in the assembly. This is because of the fact that a colloid can be repelled due to thermal gradient from the optical potential if the distance between two colloids become very less. Smaller size colloids can also form such a self-evolving colloidal matter, as shown for iron oxide infused polystyrene colloids of diameter $0.536$ $\mu$m in supporting information S16.

\section{Conclusion}

In conclusion, we report an experimental study of single defocused laser beam enabled evolving active colloidal matter. The optothermal interaction aided by the optical trapping of the active colloids in both focused and defocused laser trap enables self-thermophoretic response of the trapped colloids as well as a resultant temperature distribution. This serves as an environmental cue for the other freely diffusing ACs and lead to thermophoresis enabled long-range attraction and short-range repulsion between them. The colloids form an ordered assembly under the counteracting effect of optical gradient potential enabled attraction and short-range repulsion due to thermal gradient. The resultant colloidal matter self-evolves as a new active colloid joins the assembly. We further report input polarization state as a parameter for modulation of the self-thermophoretic motion of the ACs as well as structural orientation of the colloids in defocused laser spot. This observation is further understood by investigation of the focal electric field intensity distribution under such illumination configuration. Through this method, the number of ACs in the colloidal matter can be modulated by changing the extent of laser defocusing. Thus, our study achieves dynamic pattern formation of active colloids in a simple defocused optical field. The study can serve as a model system for understanding collective dynamics of active matter systems as well as harnessed as a re-configurable microscopic engine.


\begin{acknowledgement}
The authors acknowledge financial support from Air Force Research Laboratory grant (FA2386-18-1-4118 R\&D 18IOA118) and Swarnajayanti fellowship grant (DST/SJF/PSA-02/2017-18).
DP and RC thank Sumit Roy, Vanshika Jain, Prof. Pramod Pillai and Prof. M. Jayakannan for helping in characterization of the colloids. DP thanks Deepak K. Sharma, Ashutosh Shukla and Chaudhary Eksha Rani for fruitful discussions. The authors thank the anonymous reviewers for providing suggestion which have improved the quality of the report.
\end{acknowledgement}

\begin{suppinfo}

Supporting information containing following information is available with the manuscript. S1: Scanning electron micrograph  and ultraviolet-visible spectra of PS AC, S2: Diffusion Constant measurement, S3: Power dependence of optical trap stiffnes of PS AC, S4: Optical field distribution calculation, S5: Non-linear power dependence of off-center position, S6: Estimation of trapping force and drag velocity, S7: Temperature measurement using 5CB, S8: Temperature distribution calculation, S9: Power dependence of hovering distance, S10: Spatial distribution of optical trapping potential and gradient force, S11: Effective medium parameter calculation for PS AC, S12: Optical gradient potential and force on PS AC in defocused illumination, S13: Effect of Optical binding, S14: Simulation for Buoyancy driven convection, S15: Calculation of Soret coefficient, S16: Assembly of PS AC of diameter 0.536 $\mu$m

Supplementary videos (\href{https://youtube.com/playlist?list=PLVIRTkGrtbruGG2_wWC5CeA-4t_POOC_-}{Link}): SI Video 1: Self-evolving assembly of PS ACs, SI Video 2: Trapping of passive PS colloids, SI Video 3: Thermo-osmotic slipping of PS AC, SI Video 4: Polarization dependence of slipping direction, SI Video 5: Thermophoretic hovering, SI Video 6: 2 PS AC in defocused trap, SI Video 7: Higher number colloids. The videos are available: \url{https://youtube.com/playlist?list=PLVIRTkGrtbruGG2_wWC5CeA-4t_POOC_-}

\end{suppinfo}

\bibliography{PS_assembly.bib}

\newpage


\section*{Supplementary material}

# Supporting Information

# Optothermal evolution of active colloidal matter in defocused laser trap

Diptabrata Paul[*,†,‡], Rahul Chand[*,‡], and G V Pavan Kumar[*,†]

[*]Department of Physics, Indian Institute of Science Education and Research (IISER), Pune 411008, India
[‡]These authors contributed equally to this work
[†]email: diptarbata.paul@students.iiserpune.ac.in, pavan@iiserpune.ac.in

# Contents

# S1 Scanning electron micrograph and ultraviolet-visible spectra of PS AC

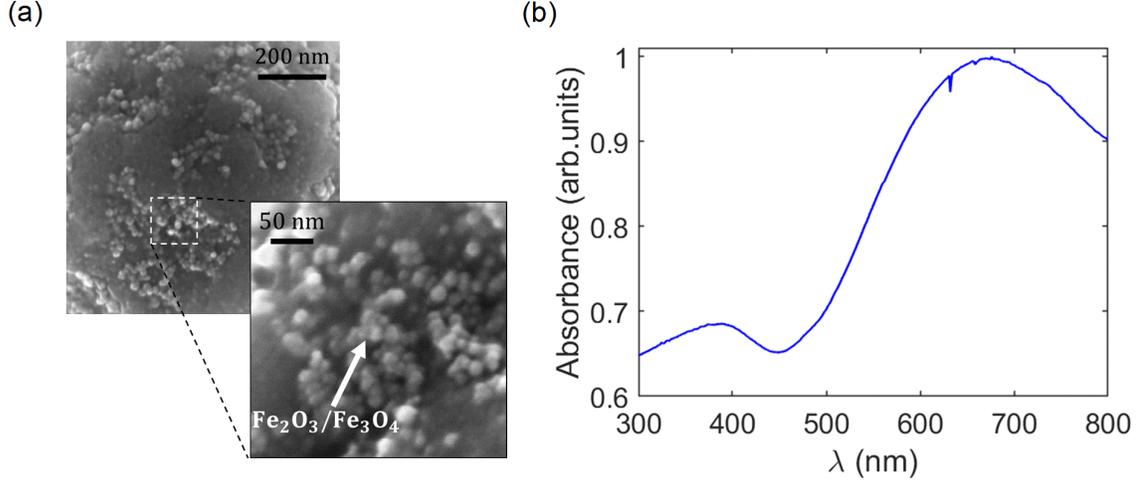

Fig. S1: (a) A scanning electron micrograph (SEM) of a PS AC having diameter 1.31 $\mu$m. A zoomed in section shows $Fe_2O_3/Fe_3O_4$ nanoparticles having diameter $\approx$ 15 nm. (b) The UV-Visible absorbance spectra of the PS ACs.

# S2 Diffusion Constant measurement

The bulk diffusion constant for a freely diffusing PS AC can be obtained theoretically by considering the equation $D_0 = \frac{k_B T}{\gamma_0}$ where $k_B$ is the Boltzmann Constant, T absolute temperature, and $\gamma_0 = 6\pi\eta_0 a$ is the bulk viscous coefficient, $a$ radius of PS AC and $\eta_0$ bulk viscosity of the fluid.

Near to a surface, these values (D,$\gamma$,$\eta$) differ significantly from the corresponding bulk value and a correction factor $\alpha$ has to be considered [1, 2], which is given by:

$$\alpha = \frac{\gamma}{\gamma_0} = \frac{\eta}{\eta_0} = \frac{D_0}{D} \qquad (1)$$

$$\alpha^{-1}(h) = 1 - \frac{9}{16}\left(\frac{a}{h}\right) + \frac{1}{8}\left(\frac{a}{h}\right)^3 - \frac{45}{256}\left(\frac{a}{h}\right)^4 - \frac{1}{16}\left(\frac{a}{h}\right)^5 + \ldots \qquad (2)$$

To obtain the experimental diffusion constant we have analysed two-dimensional mean squared displacement (MSD) of the freely diffusing colloids near the glass surface (Fig. S2 (a)) and fitted it with the equation $MSD(\tau) = 4D\tau$, where $\tau$ represents lagtime. The obtained diffusion constant value $D = 0.14$ $\mu m^2 s^{-1}$ varies significantly from the theoretically calculated bulk value $D_0 = 0.33$ $\mu m^2 s^{-1}$.

Thus, $\alpha^{-1} = D/D_0 = 0.42$, indicates the distance of the colloid from the surface. Comparing this value to the Fig. S2 (c), we find $h/a = 1.09$, implying that PS ACs are very close to the surface.

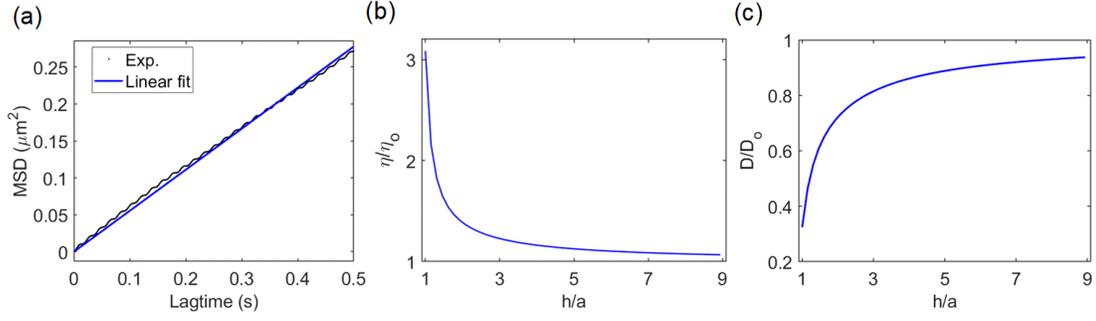

Fig. S2: Diffusion constant of PS AC. (a) Experimentally measured mean squared displacement (MSD) in two-dimension is fitted with $\mathrm{MSD}(\tau) = 4D\tau$, $\tau$ is the lag time. (b) Variation of viscosity of water ($\eta$) as a function of distance from the surface. $\eta$ near a surface deviates significantly from the bulk value $\eta_0$. Consequently, the diffusion constant (D) also deviates significantly from its bulk value ($D_0$) and its variation as a function of distance from the surface is shown in (c).

## S3  Power dependence of optical trap stiffness of PS AC

Optical trapping of a PS AC was performed with a focused laser beam at wavelength $\lambda = 532$ nm at low excitation power to minimize optical heating. The optical trap stiffness ($k$) can be obtained by analysing the position distribution of the trapped colloid, which scales linearly with laser power ($P_0$).

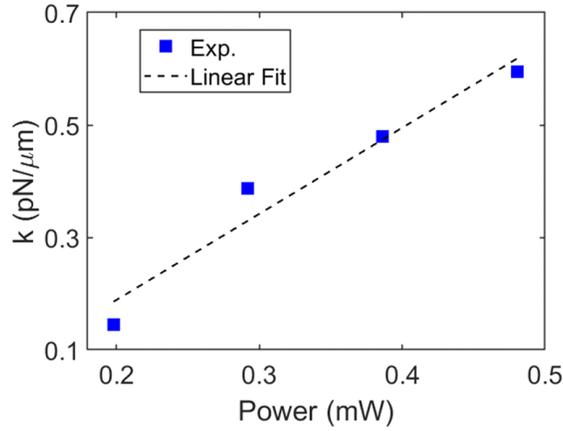

Fig. S3: Variation of the optical trap stiffness ($k$) as a function of the incident laser power ($P_0$), showing linear dependence.

The linear scaling of $k$ (in units of pN/$\mu$m) with the incident $P_0$ (in units of mW) is quantified by fitting it with the equation,

$$k = 1.53\,P_0 - 0.12, \tag{3}$$

At higher laser power illumination, the trap stiffness has to be extrapolated due to heating of the PS ACs and consequent thermo-osmotic slipping [3, 4].

# S4 Optical field distribution calculation

We have employed Debye-Wolf integral representation [5, 6, 7] to calculate the non-paraxial electric field intensity distribution at the focus of an objective lens. Let us consider an objective lens having focal length $f$ and numerical aperture $\text{NA} = n_1 \sin(\theta_{\text{max}})$, where $n_1$ is the refractive index of the medium. For an incident paraxial beam having the electric field amplitude $\mathbf{E_{in}}$ and wave vector $\mathbf{k} = k\hat{\mathbf{z}}$, the complex electric field at the focal plane of the objective lens is given by,

$$\mathbf{E}(\rho, \varphi, z) = -\frac{ikfe^{-ikf}}{2\pi} \int_0^{\theta_{\text{max}}} \int_0^{2\pi} \mathbf{E_{ref}}(\theta, \phi) e^{ikz\cos\theta} e^{ik\rho\sin\theta\cos(\phi-\varphi)} \sin\theta \, d\phi \, d\theta \tag{4a}$$

$$\mathbf{E_{ref}} = [t^s [\mathbf{E_{in}} \cdot \mathbf{n}_\phi] \mathbf{n}_\phi + t^p [\mathbf{E_{in}} \cdot \mathbf{n}_\rho] \mathbf{n}_\theta] \sqrt{\cos\theta} \tag{4b}$$

$t^s$ and $t^p$ are the transmission coefficients corresponding to s and p polarized electric fields respectively. Here $\mathbf{n}_\theta$ and $\mathbf{n}_\phi$ are unit vectors of spherical polar coordinate system and $\mathbf{n}_\rho$ and $\mathbf{n}_\phi$ are the unit vectors of cylindrical coordinate.

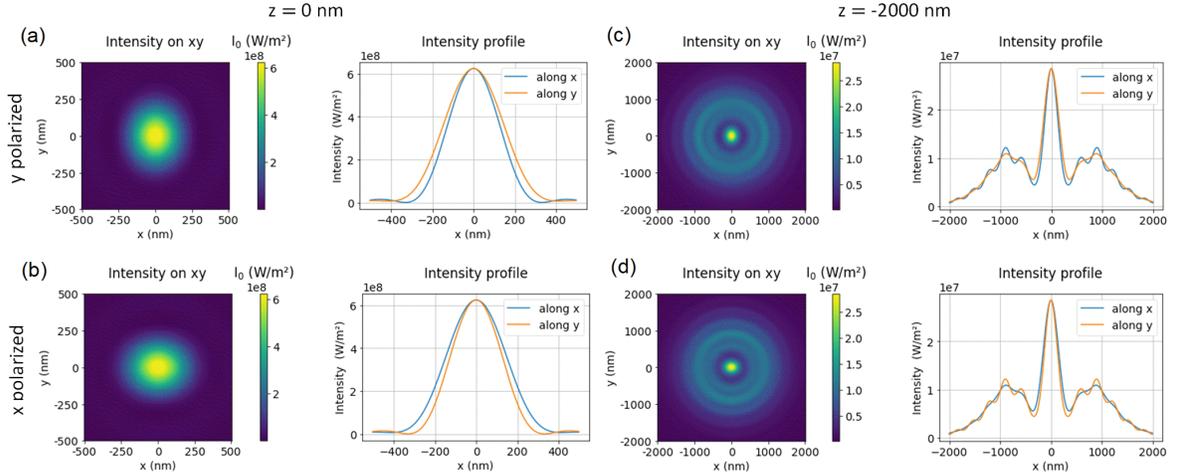

Fig. S4: Numerically calculated electric field intensity distribution and corresponding line profile along $x$ and $y$ axis for (a) focused $y$ and (b) $x$ polarized incident beam. The defocused beam intensity distribution at $z = -2000$ nm plane and the corresponding line profile along $x$ and $y$ axis for (c) $y$ polarized and (d) $x$ polarized incident beam.

Numerically calculated optical field using PyFocus [8] for focused optical field at $z = 0$ nm and the corresponding line profile along $x$ and $y$ axis for $y$ polarized and $x$ polarized incident beam is shown in Fig. S4 (a) and (b) respectively. Both these field profiles are elliptic in shape with their major axis along the polarization direction, apparent from the corresponding line profile. For defocused beam, the electric field intensity distribution is calculated at $z = -2000$ nm. The intensity distribution and the corresponding intensity profile along $x$ and $y$ axis shows that on the outer ring part, the intensity is more along $x$ axis for $y$ polarized beam (Fig. S4 (c)) and vice versa for $x$ polarized incident beam case (Fig. S4 (d)).

## S5  Non-linear power dependence of off-center position

We show here the variation of the slip distance ($\delta x$) with respect to the focused laser power ($P_0$) as extracted from the SI video 3.

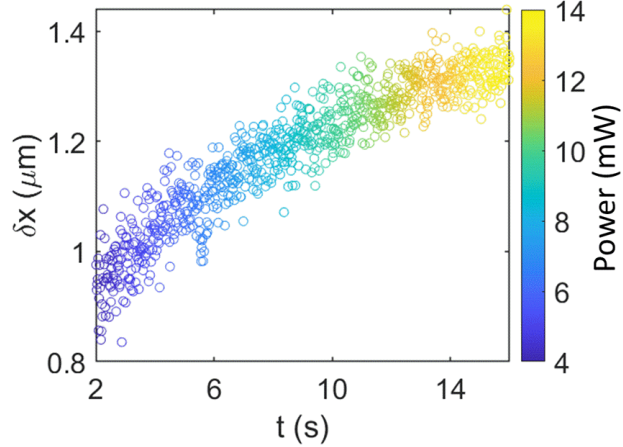

Fig. S5: Variation of the off-center self-thermophoretic slip distance ($\delta x$) with increasing laser power as extracted from SI Video 3.

In our experiments the optically trapped active colloid settles in an off-axis position where the thermal gradient induced drift force ($F_{\text{drift}} = \gamma v = -\gamma D_T \Delta T$) away from the center balances the harmonic optical force ($F_{\text{opt}} = k\delta x$) pulling the colloid towards the trap center. To analyse the non-linear dependence of the $\delta x$ of the trapped colloid with increasing laser power, we have plotted it against the estimated temperature increment of the colloid as shown in Fig. S6 (a). The temperature increment at off-center position $\delta x$ has been estimated using the equation $\Delta T^{\text{est}} = \Delta T^{\text{exp}} e^{-\delta x^2 / 2 w_0^2}$, where $\Delta T^{\text{exp}}$ is the experimentally measure temperature of an immobile centrally heated colloid and $w_0 = 0.8$ $\mu$m is the beam waist. The data was then fitted in the following ways:

**Case 1**

Considering temperature dependence of thermo-diffusion coefficient: $D_T = C(T - T^*)$ [9], We get

$$F_{\text{opt}} = F_{\text{drift}}$$

$$k\delta x = \gamma v$$

$$k\delta x = \gamma(-D_T \nabla T)$$

$$k\delta x = \gamma C (T - T^*) \left( \frac{\Delta T_0 a}{\delta x^2} \right)$$

where, $\gamma$ is the viscosity, $a$ is the radius of the particle, $k$ is the trap stiffness and $\Delta T_0$ is the increment in temperature of the trapped colloid.

$$\delta x = \left( \frac{\gamma C \Delta T_0 a}{k} \right)^{1/3} (T - T^*)^{1/3} \tag{5}$$

Since both $\Delta T_0$ and the $k$ has linear power dependence, the above equation reduces to the form:

$$\delta x = A (T - T^*)^{1/3} \tag{6}$$

Fig. S6 (b) shows the corresponding fitted plot with the equation

$$\delta x = A(T - T^*)^{1/3} + B. \tag{7}$$

Through the fit we obtain the fitting parameter $T^* = 307.7$ K.

**Case 2**

Considering temperature dependence of Soret coefficient: $S_T = S_T^\infty \left(1 - e^{(T^*-T)/T_f}\right)$ [9, 10, 11]

$$F_{opt} = F_{drift}$$

$$k\delta x = -\gamma D_T \nabla T = -\gamma S_T D \left(-\frac{\Delta T_0 a}{\delta x^2}\right)$$

$$k\delta x^3 = \gamma D S_T \Delta T_0 a = \gamma \left(\frac{k_B T}{\gamma}\right) \Delta T_0 a S_T$$

$$\delta x = \left(\frac{k_B \Delta T_0 a S_T^\infty}{k}\right)^{1/3} \left(T\left(1 - e^{(T^*-T)/T_f}\right)\right)^{1/3} \tag{8}$$

where, $k_B$ is Boltzmann constant, $a$ is the radius of the particle, $k$ is the trap stiffness and $\Delta T_0$ is the increment in temperature of the trapped particle. $S_T^\infty$ is high temperature limit of the $S_T$ and $T^*$ and $T_f$ are Soret inversion temperature and fitting parameter respectively. As before, since both $\Delta T_0$ and the $k$ has linear power dependence, the above equation reduces to the form:

$$\delta x = A \left(T\left(1 - e^{(T^*-T)/T_f}\right)\right)^{1/3} \tag{9}$$

Fitting the data in Fig. S6 (a) with the equation yields $T^* = 304.9$ K and $T_f = 4.173$ K, shown in Fig. S6 (c).

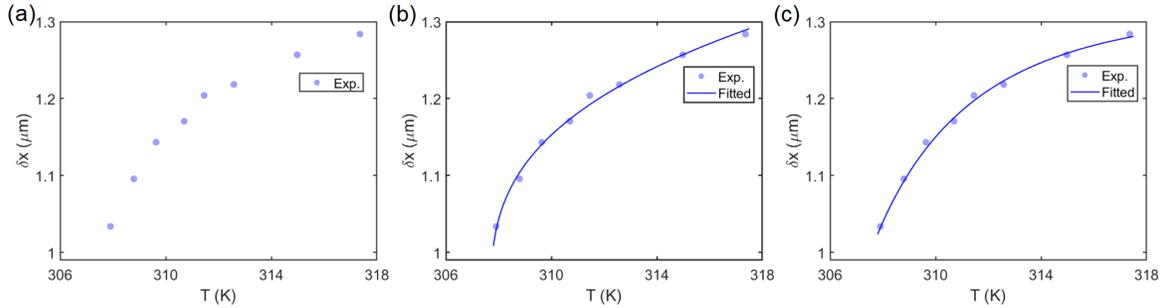

Fig. S6: Analysis of non-linear slipping distance. (a) The off-center distance of the trapped colloid as a function of estimated temperature increment. (b) The data is fitted with equation (7) from which we obtain $T^* = 307.7$ K. (c) The same data is fitted with equation (9) and obtained the fitting parameter are $T^* = 304.9$ K and $T_f = 4.173$ K.

Both cases render obtaining the value of $T^*$ similar to what reported in the literature for various colloids [9, 10, 11]. Thus, the temperature dependence of the thermo-diffusion coefficient and the Soret coefficient can offer possible explanation for the non-linear dependence of the off-center position of the trapped colloid with increasing laser power.

# S6 Estimation of trapping force and drag velocity

The trapping state due to focused optical beam has been analysed by calculating the optical force and the corresponding thermally induced drift velocity. The optical force as shown in Fig. S7 (a) is calculated by $F_{\text{opt}} = k\delta x$, where $\delta x$ represents the off-center position the colloid due to optical heating and $k$ is the trapping stiffness at corresponding power estimated by the equation $k(P_0)\,[\text{pN}/\mu\text{m}] = 1.53\,P_0\,[\text{mW}] - 0.12$ (see section S3). The extent of heating of the colloid at the off-center position $\delta x$ is estimated by: $\Delta T^{\text{est}} = \Delta T^{\text{exp}} e^{-\delta x^2/2w_0^2}$, where $w_0 = 0.8$ $\mu$m is the beam waist and $\Delta T^{\text{exp}}$ is the experimentally measured temperature of an immobile centrally heated colloid.

The drift velocity generate due to optical heating of the colloid is given by: $v = k\delta x/\gamma$. Fig. S7 (b) show the corresponding increase of drift velocity with estimated increment of the temperature of the colloid $T = \Delta T^{\text{est}} + T_s$. Here we have considered the temperature dependence of the dynamic viscosity ($\eta$) as shown in Fig. S7 (c) and corrected for the distance from the bottom glass boundary (see section S2).

$$\eta(T) = \alpha\left(1.380 - 2.122\cdot 10^{-2}\,T + 1.360\cdot 10^{-4}\,T^2 - 4.645\cdot 10^{-7}\,T^3\right) \tag{10}$$

where $\alpha = 1/0.42 = 2.38$ is the correction factor for distance.

The approximate temperature gradient can be calculated by: $|\nabla T(\delta x)| = \Delta T^{\text{exp}} \frac{a}{\delta x^2}$ which is of the order $\sim 10^7$ K m$^{-1}$.

The thermophoretic velocity of the colloid can be represented as:

$$v = \frac{2}{3}\chi\beta\frac{\nabla T}{T} \tag{11}$$

where, $\chi$ is the thermo-osmotic coefficient and $\beta = 2\,\kappa_{\text{w}}/(2\,\kappa_{\text{w}} + \kappa_{\text{PS}}) \approx 1.3$ ($\kappa_{\text{w}} = 0.6$ W m$^{-1}$ K$^{-1}$ and $\kappa_{\text{PS}} = 0.186$ W m$^{-1}$ K$^{-1}$), accounts for the difference in thermal conductivities. Considering $v \sim 5\cdot 10^2$ $\mu$m s$^{-1}$ and $T \sim 300$ K, the ratio $v/(\frac{\nabla T}{T}) = \frac{2}{3}\chi\beta \approx 1.50\cdot 10^{-8}$ m$^2$s$^{-1}$. This leads to rough estimation of the parameter $\chi \sim 1.73\cdot 10^{-8}$ m$^2$s$^{-1}$. With the thermophoretic velocity defined as $v = -D_T\nabla T$, the order of magnitude of approximate value of thermo-diffusion coefficient turns out to be $D_T = 2\chi/3T \approx 38.5$ $\mu$m$^2$ K$^{-1}$ s$^{-1}$.

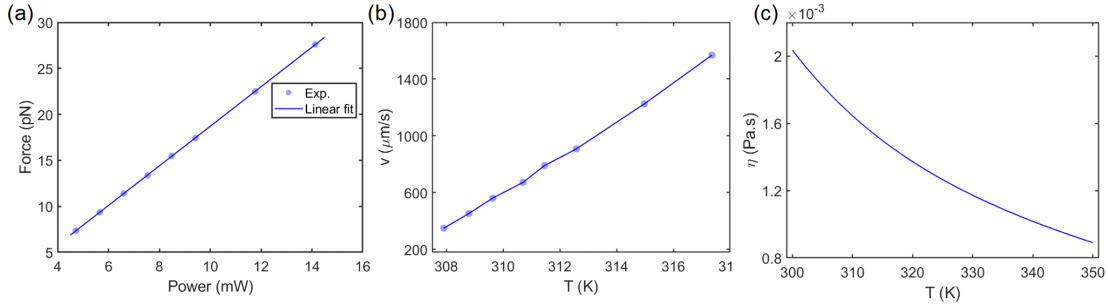

Fig. S7: Analysis of optical trapping state. (a) Optical gradient force as a function of laser power considering the off-center distance of the colloid and power dependence of the trapping stiffness. (b) The drift velocity as a function of the estimated maximum temperature of the colloid is obtained by diving the optical gradient force by the bulk viscous coefficient ($\gamma$). (c) Temperature dependence of the dynamic viscosity ($\eta$).

# S7 Temperature measurement using 5CB

The maximum surface temperature of an optically heated PS AC is obtained by studying the nematic to isotropic phase transition of 5CB liquid crystal [12, 13]. The phase transition in 5CB liquid crystal occurs upon increase of temperature beyond the phase transition temperature [12] $T_{pt} = 35°$ C$= 308$ K (Fig. S8 (a)).

An optically heated PS AC sets up a temperature distribution in the surrounding region which decays as $r^{-1}$. Thus, a heated PS AC having maximum increment of surface temperature $\Delta T_0$, would lead to a temperature distribution given by, $\Delta T(r) = \Delta T_0 \frac{a}{r}$ [14], where $a$ is the radius of the colloid. Due to the difference of refractive index of nematic and isotropic phase of the 5CB, the boundary of phase transition can be clearly visible in bright field as shown in Fig. S8 (b).

Measuring the radius of the phase transition boundary (R) the surface temperature of the heated PS AC is obtained from the equation,

$$\Delta T_0 = (T_{pt} - 298K)\frac{R-a}{a}. \tag{12}$$

The temperature increment on the surface of PS AC due to laser absorption is inversely proportional to the thermal conductivity of the surrounding medium ($\kappa$).

So, by measuring the temperature increment in 5CB medium we can calculate the temperature increment in the aqueous medium (Fig. S8 (c)) using the equation:

$$\Delta T_0^{\text{water}} = \Delta T_0^{\text{5CB}} \frac{\kappa_{\text{5CB}}}{\kappa_{\text{water}}}. \tag{13}$$

where, $\kappa_{\text{water}} = 0.60$ Wm$^{-1}$K$^{-1}$ and $\kappa_{\text{5CB}} = 0.15$ Wm$^{-1}$K$^{-1}$.

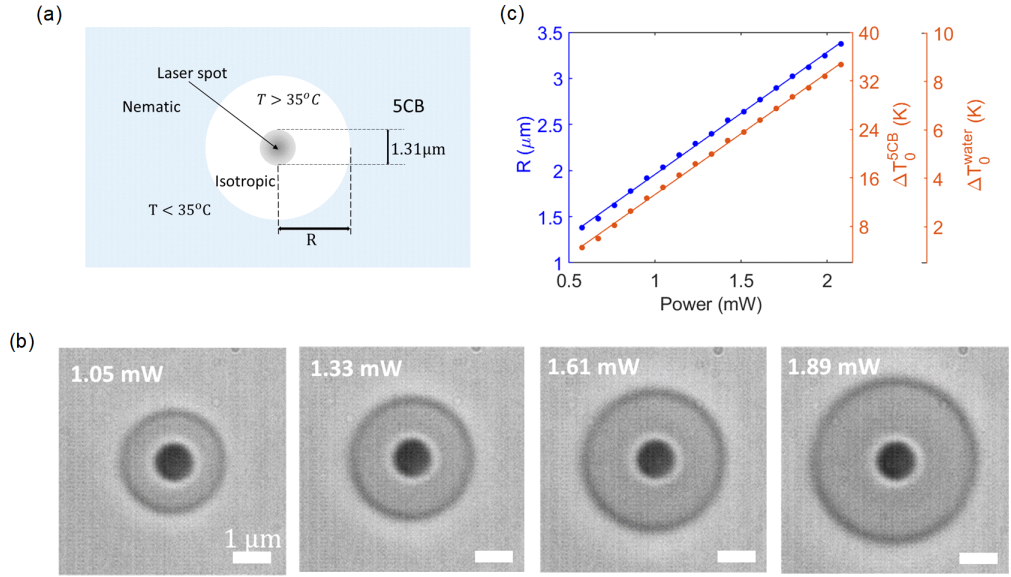

Fig. S8: Measurement of surface temperature using 5CB liquid crystal. (a) The measurement technique utilises an immobilised PS AC heated centrally with 532 nm laser and the resulting phase transition due to temperature distribution in the surrounding region.(b) The bright field optical image of 5CB phase transition for incremental incident laser power shows increase of isotropic region. (c) The maximum surface temperature of the PS AC increases linearly as the laser power is increased ($\Delta T[K] = 5.030\,P_0[mW] - 0.884$).

In case of our experimental defocused illumination the maximum surface temperature increment of the colloids can be obtained by the equation, $\Delta T[K] = 0.801\,P_0[mW] - 1.658$.

## S8  Temperature distribution calculation

Considering the surface temperature increment of a heated PS AC to be $T_0$, the temperature distribution in the surrounding region ($x$-$y$ plane) can be obtained by [14]:

$$\Delta T(r) = \begin{cases} T_0 \frac{a}{r}, & \text{for } r \geq a. \\ T_0, & \text{for } r \leq a. \end{cases} \quad (14)$$

where, $r$ is distance from the heated colloid center (see schematic Fig. S9 (a)).

This approach can be extrapolated to find the approximate temperature distribution due to multiple heated colloids through,

$$\Delta T(r) = \sum_n \Delta T_n(r_n), \quad (15)$$

where, the subscript $n$ denotes the $n^{\text{th}}$ heated colloid in the assembly and $\Delta T_n(r_n)$ is the temperature distribution sets up by the $n^{\text{th}}$ PS AC.

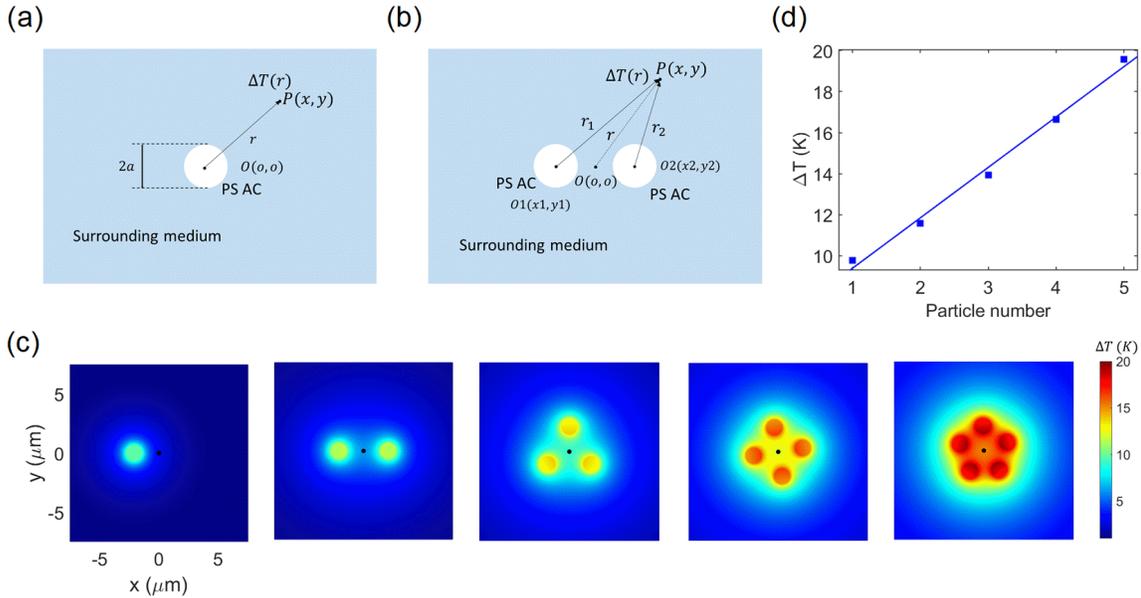

Fig. S9: Temperature distribution in the medium. (a) Schematic for calculation of temperature distribution ($\Delta T$) in the surrounding region due to a single heated PS AC (radius $a$), whose surface temperature increment can be measured by using the protocol described in section S7. (b) Schematic for calculation of temperature distribution in the $x$-$y$ plane due to two equally heated colloids. (c) The temperature distribution in the $x$-$y$ plane due to assembly of multiple colloids arranged in structures as shown in main manuscript Fig. 5. The calculations are extrapolated from measured surface increment of a single colloid, $T_0 = 9.80$ K. The central black spot indicates the center of defocused laser spot. (d) The maximum temperature of the assembly increases with number of colloids.

For example, the temperature distribution for two PS ACs can be calculated as (see schematic Fig. S9 (b)),

$$\Delta T(r) = \Delta T_1(r_1) + \Delta T_2(r_2). \quad (16)$$

where, $r_1$ and $r_2$ are the distances from the two colloids respectively.

Considering the PS ACs to be equally heated under the same optical field illumination then the temperature distribution for multiple colloids can be calculated by measuring the maximum surface temperature increment of one. Fig. S9 (c) shows the temperature distribution for the assembly of multiple colloids in the focal plane as shown in SI Video 1 and main manuscript Fig.

5. The corresponding maximum temperature increment of the system is given in Fig. S9 (d). The temperature increment values represent an approximation of the actual experimental temperature increment.

## S9 Power dependence of hovering distance

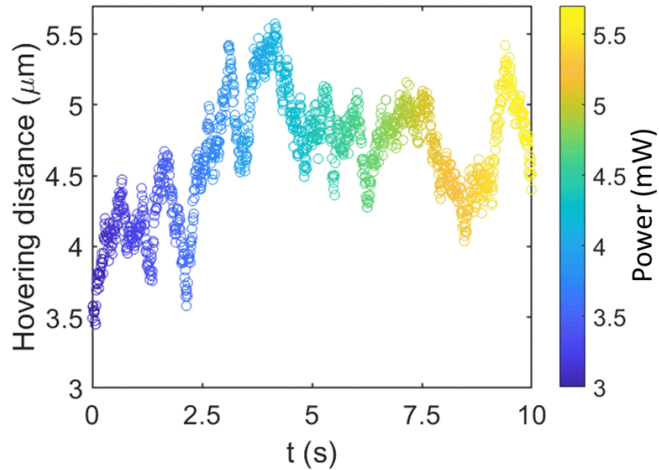

Fig. S10: Variation of the distance between a trapped PS AC and a hovering PS AC with increasing laser power extracted from SI Video 5.

## S10 Spatial distribution of optical trapping potential and gradient force

Fig. S11 (a) shows the intensity profile of the illuminated optical field. To extract the profile of the optical gradient potential and force in the sample plane we scan an immobile PS AC in 5CB medium along a line in the optical field (shown by black dashed line). The temperature increment of the colloid is dependent on the spatial profile of the optical field. Fig. S11 (b) shows the generation of isotropic phase of 5CB due to increment of surface temperature of the PS AC in three different points as indicated in Fig. S11 (a). The temperature profile along a line parallel to $x$ axis (as shown by black dashed line in Fig. S11 (a)) is shown in Fig. S11 (c). The increment in temperature is directly proportional to the intensity profile, as the case with trapping potential and the gradient force.

Since, both the temperature increment and the optical trapping potential is proportional to the intensity profile of the incident optical field, the temperature profile also gives us a qualitative measure of the optical gradient potential. Fig. S11 (d) shows a qualitative profile of optical trapping potential (in blue fitted with double gaussian function) and the corresponding optical gradient force profile (in red). The approximate trapping position is at the secondary minima of the optical potential and have been indicated by the grey dashed lines.

The quantitative measure have been obtained by trapping passive polystyrene particles having diameter 1.01 $\mu$m in the defocused optical field at laser power $P_0 = 14$ mW (SI Video 2) and resulting trapping characteristics.

The passive colloids get trapped in the optical field in two regions - central region and the peripheral ring region as shown in Fig. S12 (a). The measured radial trapping potential and the

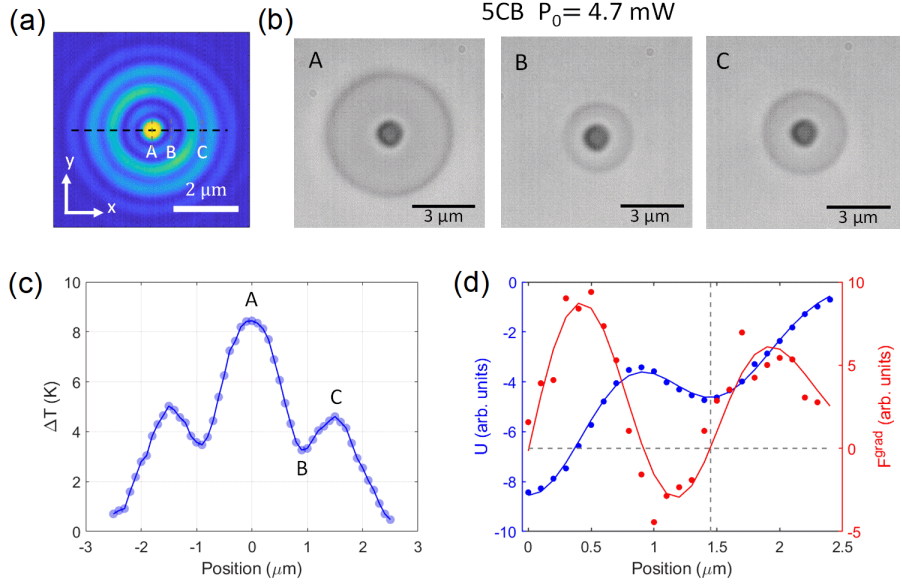

Fig. S11: Estimation of optical gradient potential and force profile. (a) Incident optical field intensity profile. (b) Bright field optical image of 5CB phase transition due to heating of an immobile PS AC placed in different points of the optical field at laser power $P_0 = 4.7$ mW. (c) Extracted temperature profile along a dashed black line in (a). (d) The corresponding optical potential and the optical gradient force line profile.

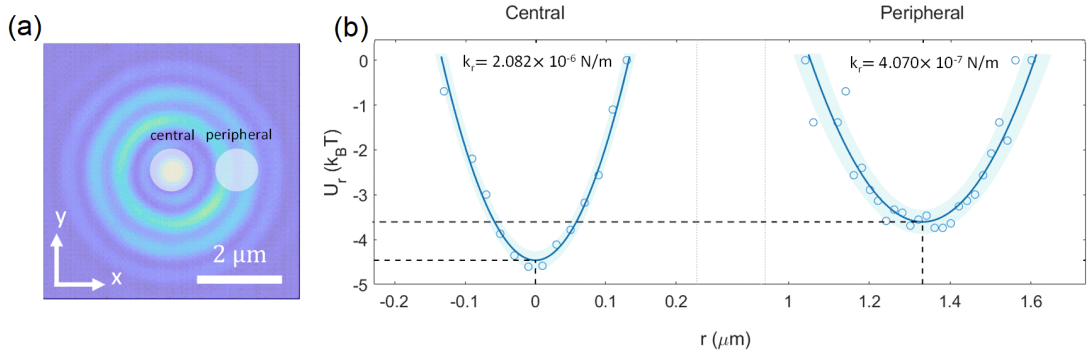

Fig. S12: Estimation of radial trapping potential for defocused illumination. (a) The passive PS colloids gets trapped in two regions of the optical field - in the central part and in the peripheral ring region. (b) The corresponding radial trapping potential. The trap stiffness is more for trapped colloid in the central region and the potential has a minimum $\approx -4.4$ $k_B T$ and trap stiffness $k_r \approx 2.082$ pN/$\mu$m. The peripheral region has minima $\approx -3.6$ $k_B T$ and trap stiffness $k_r \approx 0.407$ pN/$\mu$m.

trap stiffness is more for the centrally trapped colloid with respect to the peripherally trapped colloid as can be seen from Fig. S12 (b). For the centrally trapped colloid the minima of the trapping potential were found to be $U_r^{cent} \approx -4.4$ $k_B T$ and trap stiffness $k_r \approx 2.082$ pN/$\mu$m. On the other hand, trapping potential for the trapped colloid in the peripheral ring region has minima $U_r^{peri} \approx -3.6$ $k_B T$ and trap stiffness $k_r \approx 0.407$ pN/$\mu$m. The force on the particle at the central spot is approximately an order of magnitude higher than that in the peripheral region.

## S11 Effective medium parameter calculation for PS AC

Considering the iron oxide inclusions distributed uniformly through out the polystyrene active colloids, we can get the estimate of effective dielectric constant ($\varepsilon_{\text{eff}}$) of PS ACs by employing Maxwell-Garnett effective medium approximation as [15],

$$\frac{\varepsilon_{\text{eff}}}{\varepsilon_{\text{PS}}} = \frac{1 + 2\phi \frac{\varepsilon_{\text{i}} - \varepsilon_{\text{PS}}}{\varepsilon_{\text{i}} + 2\varepsilon_{\text{PS}}}}{1 - \phi \frac{\varepsilon_{\text{i}} - \varepsilon_{\text{PS}}}{\varepsilon_{\text{i}} + 2\varepsilon_{\text{PS}}}} \tag{17}$$

where $\varepsilon_i$ and $\varepsilon_{\text{PS}}$ are the dielectric constant of the iron oxide inclusions and polystyrene respectively. The volume fraction of the iron oxide inclusions ($Fe_2O_3/Fe_3O_3$) is obtained by assuming a thin shell (thickness $t$) of iron oxide around a passive dielectric polystyrene sphere, the combination of both results in a sphere of radius $R$. The 30 % mass fraction of the of iron oxide inclusion in the PS ACs results in volume fraction $\phi = 7.7\,\%$. Assuming $1:1$ ratio of $Fe_2O_3 : Fe_3O_3$, the effective dielectric constant turns out to be $\varepsilon_{\text{eff}} = 2.8145 + 0.0567i$, which leads to the effective refractive index $n_{\text{eff}} = \sqrt{\varepsilon_{\text{eff}}} = 1.6777 + 0.0169i$.

Similarly to get the effective thermal conductivity ($\kappa_{eff}$) we have used Maxwell's formula as [16],

$$\frac{\kappa_{\text{eff}}}{\kappa_{\text{PS}}} = 1 + \frac{3\phi}{\frac{\kappa_{\text{i}} + 2\kappa_{\text{PS}}}{\kappa_{\text{i}} - 2\kappa_{\text{PS}}} - \phi}. \tag{18}$$

Where $\kappa_{\text{i}}$ and $\kappa_{\text{PS}}$ are the thermal conductivity of the inclusion and the polystyrene media. We get effective thermal conductivity as $\kappa_{\text{eff}} = 0.1861$.

Following are the medium property used for the calculations

| Material | Polystyrene | $Fe_2O_3$ | $Fe_3O_4$ | Inclusion |
|---|---|---|---|---|
| $\varepsilon$ | 2.5546 | $10.3810 + 3.6546i$ | $5.4933 + 0.4344i$ | $7.9371 + 2.0445i$ |
| $\kappa\,(\text{W}\,\text{m}^{-1}\,\text{K}^{-1})$ | 0.15 | 7 | 17.65 | 12.33 |

Table S1: Material properties used for the calculation of effective parameters.

## S12 Optical gradient potential and force on PS AC in defocused illumination

We have calculated the optical gradient potential and optical gradient force on PS AC due to defocused optical beam by evaluating the trap stiffness from the time series data (SI Video 1) at laser power $P_0=14.13$ mW. 450 frames have been considered in the calculation during the time frame where two colloids are oriented along the $x$ axis due to the $y$-polarization of the illumination beam.

The optical gradient potential is higher along the $x$ direction with respect to the $y$ direction as shown in Fig. S13. Consequently, the trap stiffness which quantifies the optical gradient force, $k_x \approx 0.447$ pN/$\mu$m is more with respect to $k_y \approx 0.035$ pN/$\mu$m. The variation of the values can be understood by considering their orientation in the optical field.

We have numerically calculated the optical defocused field and evaluated the optical gradient force for $x$ and $y$ polarized incident field using PyFocus [8]. For a spherical colloid of radius $R \leq \lambda$ under the influence of an optical field, the time averaged radiation force can be given by:

$$\langle \mathbf{F} \rangle = \frac{1}{4}\epsilon_0 \epsilon_r \text{Re}(\alpha) \nabla |\mathbf{E}|^2 + \frac{1}{2}\varepsilon_0 \varepsilon_r \text{Im}(\alpha) \text{Im}\left(\sum_n \mathbf{E}_n^* \nabla \mathbf{E}_n\right) \tag{19}$$

where $\alpha(\omega)$ represents the complex polarizability of the colloid and $n = x, y, z$. The gradient force component can be represented as,

$$\langle \mathbf{F}^{\text{grad}} \rangle = \frac{1}{4}\epsilon_0 \epsilon_r \text{Re}(\alpha) \nabla |\mathbf{E}|^2 \tag{20}$$

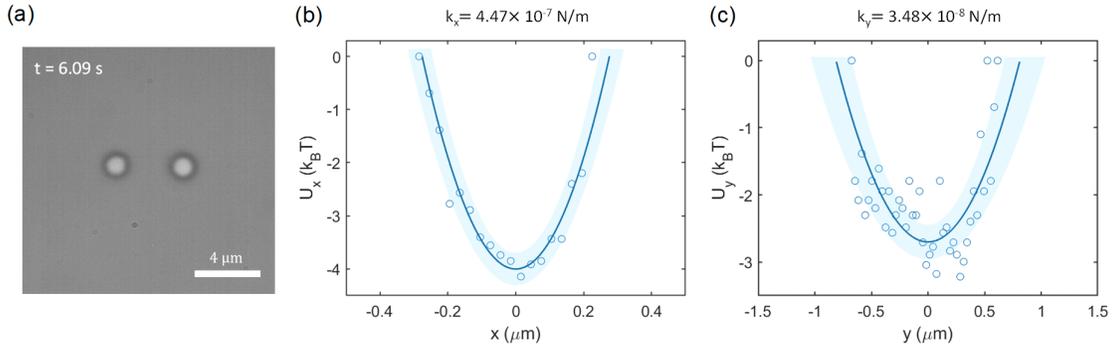

Fig. S13: Calculation of optical gradient potential and force on active colloids for defocused illumination. (a) assembly of two colloids oriented along $x$ axis in the defocused field. Optical gradient potential along (b) $x$ axis and (c) $y$ axis on the active colloids due to the $y$ polarized defocused illumination. The potential and the trap are much stiffer along $x$ axis compared to the $y$ direction.

Using this equation, the order of magnitude of optical gradient force on the active colloids can be calculated. For active colloids of radius $R = 0.65\,\mu m$, the complex polarizability is given by:

$$\alpha(\omega) = \frac{1 - (1/10)(\varepsilon_2(\omega) + \varepsilon_1)\xi^2}{(1/3 + \varepsilon_1/(\varepsilon_2(\omega) - \varepsilon_1)) - (1/30)(\varepsilon_2(\omega) + 10\varepsilon_1)\xi^2 - i 4\pi^2 \varepsilon_1^{3/2} V/(3\lambda^3)} \quad (21)$$

where $\xi = 2\pi R/\lambda$ is the size parameter, V is the volume of the colloids, $\varepsilon_2(\omega)$ and $\varepsilon_1$ are the electric permittivity of the polystyrene active colloids and the surrounding water media respectively. The electric permittivity of the iron oxide infused colloid has been modelled using Maxwell-Garnett effective medium theory (see section S11) [15].

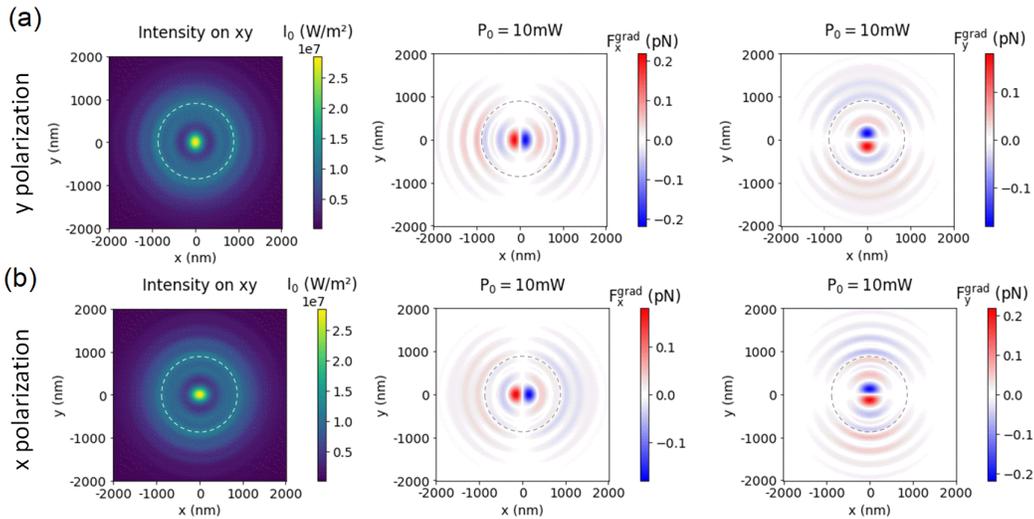

Fig. S14: Numerical calculation of optical gradient force. Incident optical field ($I_0$), optical gradient force along $x$ direction ($F_x^{\mathrm{grad}}$) and $y$ direction ($F_y^{\mathrm{grad}}$) for (a) $y$ polarized beam and (b) $x$ polarized incident beam at power $P_0 = 10$ mW. The white dashed circle in first column and grey dashed circles in second and third column indicates the approximate radial trapping position of colloid.

Fig. S14 shows the numerically calculated optical force on the active colloids at laser power $P_0 = 10$ mW. The approximate position of trapping of the active colloid is indicated by white dashed circle in the first column and grey dashed circles in the second and third column. The optical forces

are slightly higher perpendicular to the polarization direction. Although these are approximate calculation, the order of magnitudes matches well with the experimentally obtained trap stiffness.

## S13  Effect of Optical binding

To evaluate the effect of optical binding we have used COMSOL 5.1 and employed a cuboidal geometry ($8\,\mu\text{m} \times 8\,\mu\text{m} \times 6\,\mu\text{m}$) as shown in Fig. S15 (a). A defocused Gaussian optical field at wavelength $\lambda = 532$ nm and with polarization along $y$ direction was introduced using a port as shown by the green envelope and arrow. Iron oxide infused colloids were modelled as uniform spheres having diameter 1.3 with refractive index ($n = 1.6777 + i0.0169$) obtained by considering Maxwell-Garnett effective medium theory (see section S11) [15]. Scattering boundary conditions were applied to appropriate boundaries to avoid spurious reflections. The medium was modelled as homogeneous with refractive index set to that of water ($n_{\text{med}} = 1.33$). The entire geometry has been meshed using a minimum tetrahedral meshing size of $\sim 150$ nm.

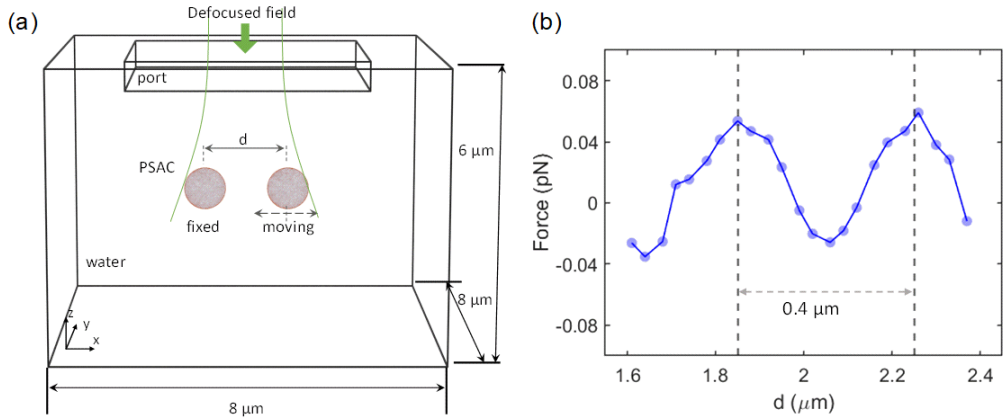

Fig. S15: Numerical calculation of optical scattering force. (a) Schematic of the geometry employed for calculation of optical scattering force. (b) Evaluated optical scattering force using Maxwell stress tensor at the fixed particle as a function of position of the second moving particle.

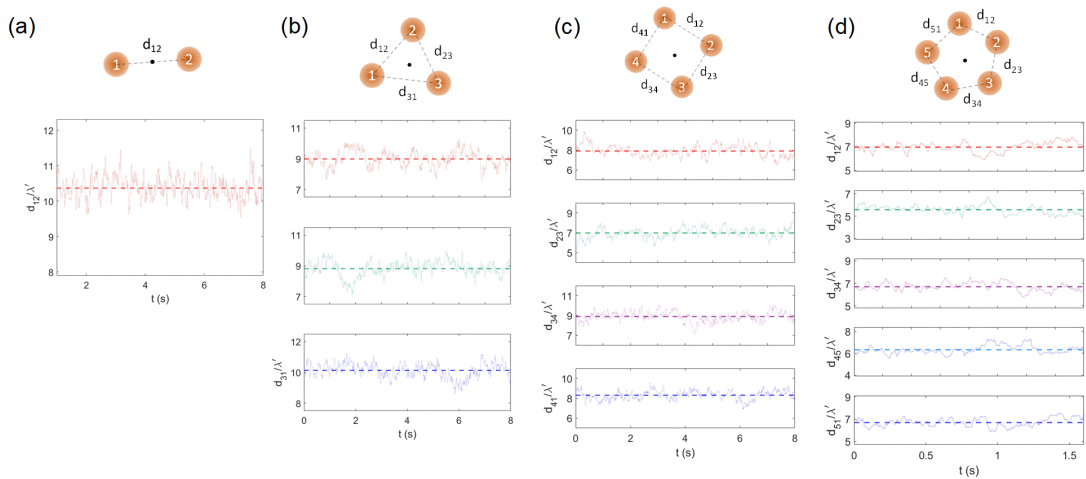

Fig. S16: Analysis of interparticle distances as multiple of $\lambda' = 0.4\ \mu$m of the colloidal assembly for (a) 2 particles, (b) 3 particles, (c) 4 particles and (d) 5 particles.

S14

The optical scattering force have been calculated by employing the Maxwell stress tensor formulation of the wave-optics module. A colloid was fixed at one position, and the scattering force on the particle was calculated as a function of position of the second moving colloid as shown in Fig. S15 (a). Fig. S15 (b) shows the numerically calculated optical scattering force. The force oscillates as function of position of the moving colloids with period $\lambda' = \lambda/n_\mathrm{med} = 400$ nm. The oscillations are characteristics of optical binding force and indicative of the scattering force on adjacent particles which are in the order of $\sim 0.01$ pN, one order smaller than the optical gradient force on the colloids. This represents the approximate scattering force on the colloids as implementation of strong focusing and resultant vectorial optical field is difficult to achieve in COMSOL.

Further, we have analysed the trajectory of the colloidal assembly and calculated the time series of interparticle distance as $d/\lambda'$, shown in Fig. S16. For colloidal pair, the optical binding effect is minimal as the interparticle distance is not always integer multiple of $\lambda'$. But for higher number of particles in the assembly, a weaker effect of optical binding can be inferred as interparticle distances are roughly integer multiples of $\lambda'$.

## S14 Simulation for Buoyancy driven convection

To enumerate the effect of buoyancy-driven convection in our experimental configuration, we have used COMSOL 5.1 and combined the heat-transfer module with laminar flow module. To get an estimate we only consider a centrally heated colloid and employ a 2D axisymmetric model geometry for the calculation as shown in Fig. S17. This approach is taken instead of a 3D simulation since numerical calculations of such nature are computationally exhaustive.

The outer boundary layer was set to $T_0 = 298$ K. The colloid (PS AC) was defined as the heat source with the equation:

$$Q(r,z) = \frac{2P_0}{\pi w_0^2} \alpha e^{-\alpha z} e^{-2r^2/w_0^2} \qquad (22)$$

where, $P_0$ is the incident laser power, $w_0 = 0.8$ $\mu$m is beam waist and $\alpha$ is the absorption coefficient of the modelled particle at wavelength $\lambda = 532$ nm. Modelling the iron oxide infused colloid using Maxwell-Garnett effective medium theory and calculating the refractive index as ($n = 1.6777 + i0.0169$), the absorption coefficient was calculated using the equation $\alpha = 4\pi n''/\lambda' = 5.3069 \cdot 10^5$ m$^{-1}$, $\lambda' = \lambda/n_\mathrm{med}$, $n''$ denoting the imaginary part of the refractive index. A temperature dependent density $\rho(T)$, dynamic viscosity $\eta(T)$ as well as thermal conductivity $\kappa(T)$ of the water facilitates introduction of thermal convection.

Additionally, the following parameters were used:

| Materials | $\kappa$(Wm$^{-1}$K$^{-1}$) | $\rho$(Kgm$^{-3}$) | $C_p$(JKg$^{-1}$K$^{-1}$) |
|---|---|---|---|
| Glass | 1.38 | 2203 | 703 |
| Colloid | 0.1861 | 1215 | 1070 |

Table S2: Material properties used for numerical simulations.

Fig. S18 (a) shows the calculated temperature distribution. The temperature is more towards the central part of the colloid due to its lower thermal conductivity. However, experimental temperature only reflects the surface temperature. Fig. S18 (b) shows the corresponding temperature distribution along a line parallel to $x$ axis at $z = 0.65$ $\mu$m, which shows decreasing trend from the central region. The surface temperature as a function of laser power is shown in Fig. S18 (c) and matches reasonably well with the experimentally measured temperature. The relative density change is shown in Fig. S18 (d).

To calculate the thermal convection, the upper and lower glass-water interfaces as well as the colloid water interface were set with no-slip boundary conditions and the boundary on right side as open boundary as shown in Fig. S17. The resultant convective flow field and its $x$ and $z$ components

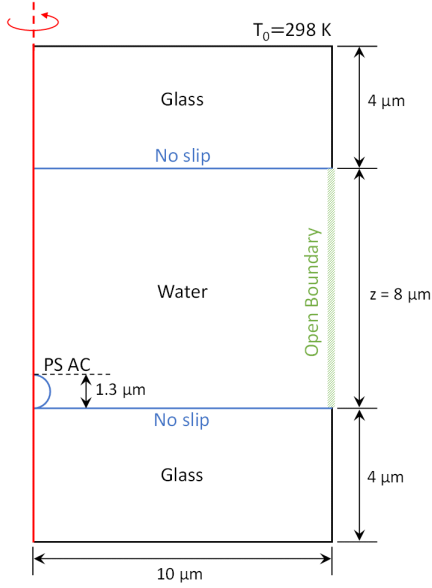

Fig. S17: Schematic of the 2D axisymmetric geometry employed to calculate the temperature and the corresponding buoyancy driven convection.

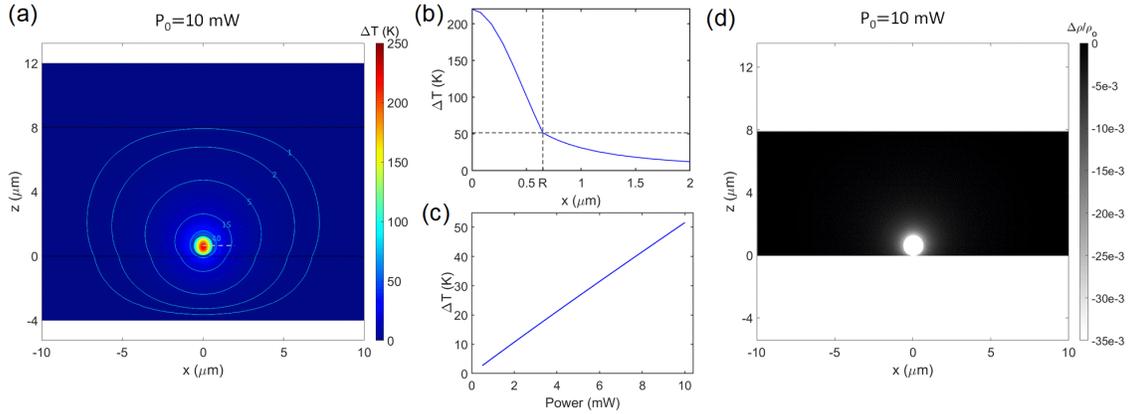

Fig. S18: Simulated temperature distribution and density distribution considering thermal convection. (a) Temperature distribution in the $xz$ plane (b) Temperature distribution along a line parallel to the $x$ axis at $z = 0.65$ $\mu$m. (c) Temperature at the surface of the colloid as function of power shows increasing linear trend. (d) The relative density change in the $xz$ plane.

are shown in Fig. S19. The calculated maximum flow velocity even for 10 mW excitation of a centrally heated colloid are of the order $\sim 10^1$ nm s$^{-1}$, which are approximately two orders of magnitude lower than the experimentally observed thermophoretic velocity. The calculated maximum velocity saturates at about 63 nm s$^{-1}$ as the sample chamber height is increased to 120 $\mu$m as shown in Fig. S20.

An assembly of heated colloid may act like an extended heat source but for our experimental configuration even for an assembly of five colloids in the defocused beam at power P$_0$ = 14 mW, the estimated temperature increment is $\approx$ 20 K (supporting information S8). Thus, the effect of buoyancy driven thermal convection can be considered minimal with respect to the thermophoretic motion of the colloids.

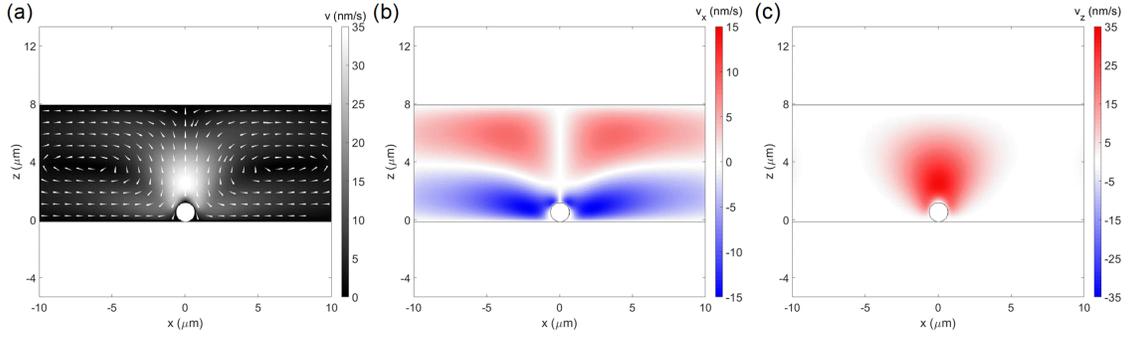

Fig. S19: Simulated flow field due to thermal convection for a laser power of $P_0 = 10$ mW at $xz$ plane. (a) Magnitude of the thermal flow field along with its direction (white arrow heads). The corresponding $x$ and $z$ component of the flow field is shown in (b) and (c) respectively.

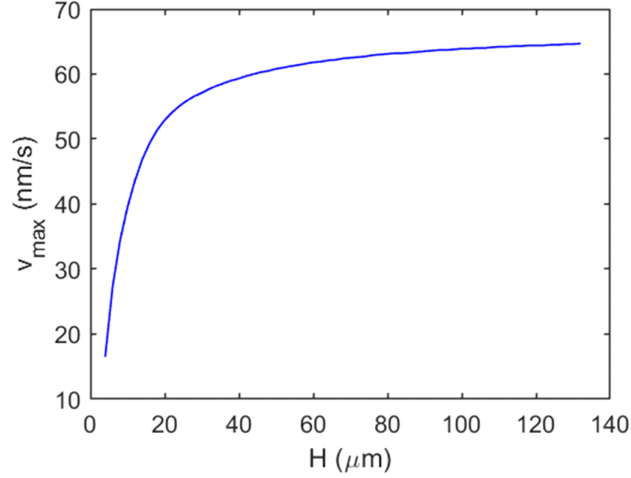

Fig. S20: Simulated maximum convective velocity $v_{max}$ at laser power $P_0 = 10$ mW as a function of chamber height H.

## S15 Calculation of Soret coefficient

From the thermophoretic velocity ($v$) of second PS AC under the influence of temperature distribution set up by the first trapped PS AC in the defocused optical field, we can obtain the thermo-diffusion coefficient ($D_T$) [4] through,

$$v = -D_T \nabla T. \quad (23)$$

Where $T = T_s + \frac{\Delta T_0 a}{r}$ is the absolute temperature, $T_s$ is the ambient temperature and $\Delta T_0$ is the maximum surface temperature of heated PS AC. Now inserting T in the equation (23) we obtain the trajectory ($r(t)$) of the incoming second PS AC as,

$$r(t) = (A + Bt)^{\frac{1}{3}}. \quad (24)$$

Where $A = r_0^3$ and $B = 3D_T \Delta T_0 a$ .

Thus, by fitting $r(t)$ vs. $t$, we can obtain $D_T$. Then, the Soret coefficient can be obtained as $S_T = \frac{D_T}{D}$. If $S_T > 0$ then the PS AC will move to the colder region and if $S_T < 0$, then the PS AC will move to the hotter region.

The trajectory $r(t)$ of the second colloid undergoing thermophoretic motion with respect to the trapped one as shown in Fig. S21 (a) is obtained by tracking their coordinates. The corresponding

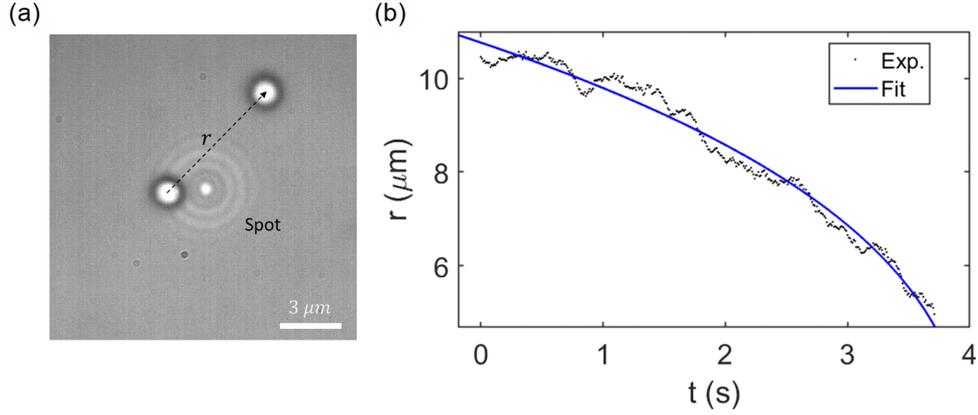

Fig. S21: Soret coefficient calculation. (a) Snapshot of the motion of a PS AC under the influence of temperature distribution set up by a trapped and heated PS AC in a defocused laser spot. (b) Trajectory of the second colloid with respect to the trapped colloid $r(t)$ is fitted with equation (24).

trajectory $r(t)$ is plotted in Fig. S21 (b). Fitting it with equation (24) we obtain the fitting parameters $A = 1248.00$ $\mu m^3$ and $B = -308.20$ $\mu m^3 s^{-1}$. Thus, the thermo-diffusion coefficient is obtained as $D_T = \frac{B}{3\Delta T_0 a} = -16.01$ $\mu m^2 K^{-1} s^{-1}$, where $\Delta T_0 = 9.80$ K and $2a = 1.31$ $\mu m$. The Soret coefficient can be obtained as $S_T = \frac{D_T}{D} = -115.16$ $K^{-1}$ ($D = 0.14$ $\mu m^2 s^{-1}$). The negative sign implies the motion towards positive thermal gradient as in our experimental case.

## S16   Assembly of PS ACs of diameter 0.536 $\mu$m

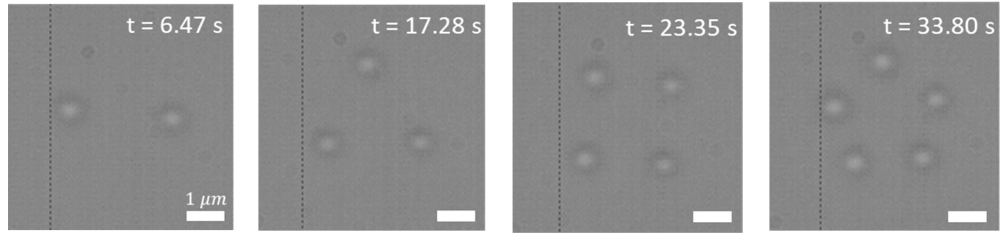

Fig. S22: Assembly formation of smaller PS ACs of diameter 0.536 $\mu$m in the defocused laser spot.



\end{document}